\documentclass[12pt,a4paper]{article}
\usepackage{amsmath}
\usepackage{amsfonts}
\usepackage{graphicx}
\usepackage{cite}
\usepackage{bbm}
\usepackage{color}
\DeclareMathOperator{\Li}{Li}

\def\({\left(}
\def\){\right)}

\newcommand{\cO}{{\cal O}}

\newcommand{\nn}{\nonumber}
\newcommand{\Eqn}[1]{&\hspace{-0.5em}#1\hspace{-0.5em}&}
\renewcommand{\[}{\begin{equation}}
\renewcommand{\]}{\end{equation}}
\newcommand{\eqb}{\begin{eqnarray}}
\newcommand{\eqe}{\end{eqnarray}}

\newcommand{\bbZ}{{\mathbb Z}}

\newcommand{\bz}{{\bar{z}}}
\newcommand{\cK}{{\cal K}}

\newcommand{\dilog}{\operatorname{Li_2}}
\newcommand{\tepsilon}{{\tilde{\epsilon}}}

\def\be {\begin{equation}}
\def\ee {\end{equation}}

\def\o{\over}

\def\comma      { \, , }
\def\period     { \, . }

\def\Im#1    { \, {\rm Im } \, #1  }
\def\Re#1    { \, {\rm Re}  \, #1  }
\def\calO  {{\cal O}}
\renewcommand{\thesection}
 {\arabic{section}.\hspace{-.5em}}
\renewcommand{\thesubsection}
 {\arabic{section}.\arabic{subsection}.\hspace{-.5em}}
\renewcommand{\thesubsubsection}
 {\arabic{section}.\arabic{subsection}.\arabic{subsubsection}.\hspace
                                                              {-.5em}}

\makeatletter
\renewcommand\section{
 \@startsection{section}{3}{\z@}%
 {-3.25ex\@plus -1ex \@minus -.2ex}%
 {1.5ex \@plus .2ex}%
 {\normalfont\normalsize\bfseries\mathversion{bold}}}
\renewcommand\subsection{
 \@startsection{subsection}{3}{\z@}%
 {-3.25ex\@plus -1ex \@minus -.2ex}%
 {1.5ex \@plus .2ex}%
 {\normalfont\normalsize\bfseries\mathversion{bold}}}
\renewcommand\subsubsection{
 \@startsection{subsubsection}{3}{\z@}%
 {-3.25ex\@plus -1ex \@minus -.2ex}%
 {1.5ex \@plus .2ex}%
 {\normalfont\normalsize\itshape}}
\makeatother

\makeatletter \@addtoreset{equation}{section} \makeatother
\renewcommand{\theequation}{\arabic{section}.\arabic{equation}}

\makeatletter
\renewcommand{\appendix}{
\renewcommand{\thesection}{Appendix \Alph{section}.\hspace{-.5em}}
\renewcommand{\thesubsection}
 {\Alph{section}.\arabic{subsection}.\hspace{-.5em}}
\renewcommand{\thesubsubsection}
 {\Alph{section}.\arabic{subsection}.\arabic{subsubsection}.\hspace
                                                            {-.5em}}
\@addtoreset{equation}{subsection}
\renewcommand{\theequation}{\Alph{section}.\arabic{equation}} 
\setcounter{section}{0}}
\makeatother

\begin{document}
%%%%%%%%%%%%%%%%%%%%%
%    cover
%%%%%%%%%%%%%%%%%%%%%
%
%%%%%%%%%%%%%%%%%%%%%%%%%%%%%%%%%%%
\def\papertitlepage{\baselineskip 3.5ex \thispagestyle{empty}}
\def\preprinumber#1#2#3#4{\hfill
\begin{minipage}{1.2in}
#1 \par\noindent #2 \par\noindent #3 \par\noindent #4
\end{minipage}}
\renewcommand{\thefootnote}{\fnsymbol{footnote}}
\newcounter{aff}
\renewcommand{\theaff}{\fnsymbol{aff}}
\newcommand{\affiliation}[1]{
 \setcounter{aff}{#1} $\rule{0em}{1.2ex}^\theaff\hspace{-.4em}$}
%%%%%%%%%%%%%%%%%%%%%%%%%%%%%%%%%%%%%%%%%%%%%%%%
%
\papertitlepage
\setcounter{page}{0}
\preprinumber{}{DESY 14-058}{TIT/HEP-636}{UTHEP-662}{}
\vskip 4ex

%\vfill
\baselineskip=4ex
\begin{center}
{\large\bf\mathversion{bold}
Quantum Wronskian approach to 
six-point gluon scattering amplitudes at strong coupling
%(tentative)
}
\end{center}
\vfill
\baselineskip=3.2ex
\begin{center}
 Yasuyuki Hatsuda\footnote[1]{\tt yasuyuki.hatsuda@desy.de}, 
 Katsushi Ito\footnote[2]{\tt ito@th.phys.titech.ac.jp},
  Yuji Satoh\footnote[3]{\tt ysatoh@het.ph.tsukuba.ac.jp} and
 Junji Suzuki \footnote[4]{\tt sjsuzuk@ipc.shizuoka.ac.jp}\\

\vskip 3ex
 \affiliation{1}
 {\it DESY Theory Group, DESY Hamburg} \\
{\it D-22603 Hamburg, Germany }\\
 %{\it }\\

\vskip 1ex
 \affiliation{2}
 {\it Department of Physics, Tokyo Institute of Technology}\\
 {\it Tokyo 152-8551, Japan}\\

\vskip 1ex
 \affiliation{3}
 {\it Institute of Physics, University of Tsukuba}\\
 {\it Ibaraki 305-8571, Japan}\\

\vskip 1ex
 \affiliation{4}
 {\it  Department of Physics, Shizuoka University}\\
 {\it Shizuoka 422-8529, Japan}

\end{center}
%%%%%%%%%%%%%%%%%%%%%%%
%\vfill
%%%%%%%%%%%%%%%%%%%%%%%
%
\baselineskip=3.3ex
\vskip 2ex
\begin{center} {\bf Abstract} \end{center}
\vskip -0.5ex
We study the six-point gluon scattering amplitudes in ${\cal N}=4$ super Yang-Mills
theory at strong coupling based on
the twisted $\bbZ_4$-symmetric integrable model. 
The lattice regularization allows us to derive
the associated thermodynamic Bethe ansatz (TBA) equations
as well as the functional relations among the Q-/T-/Y-functions.
The quantum Wronskian relation  
for the Q-/T-functions  plays an important role in determining a series of the expansion 
coefficients of the  T-/Y-functions around the UV limit, 
including the dependence on the twist parameter. 
Studying the CFT limit of the TBA equations, we 
derive the leading analytic expansion of the
remainder function for the general kinematics
around the limit where the dual Wilson loops become regular-polygonal.
We also compare the rescaled remainder functions at strong coupling 
with those at two, three  and four loops, and
find that they are close to each other 
along the trajectories parameterized by the scale parameter of the integrable model.

\vskip 3ex
\vspace*{\fill}
\noindent
June 2014
\setcounter{page}{0}
\newpage
\renewcommand{\thefootnote}{\arabic{footnote}}
\setcounter{footnote}{0}
\setcounter{section}{0}
\baselineskip = 3.5ex
\pagestyle{plain}

\baselineskip=3.5ex
%---------------------------------------------------------------------------------------------------
\section{Introduction}
%---------------------------------------------------------------------------------------------------

The gluon scattering amplitudes in 
${\cal N}=4$ super Yang-Mills 
theory are a subject of great  interest in recent years.
In the planar limit, they are dual to the null-polygonal Wilson loops whose
 segments are light-like and proportional to external gluon momenta
\cite{Alday:2007hr,Drummond:2007aua,Brandhuber:2007yx,Mason:2010yk,CaronHuot:2010ek}.
The duality implies a conformal symmetry in the dual space 
\cite{Alday:2007hr,Drummond:2006rz,Drummond:2007au,Drummond:2008vq}.
This dual conformal symmetry
strongly constrains the form of the amplitudes. 
In particular, the maximal helicity violating (MHV) amplitude
 is expressed as a  sum of 
 the  Bern, Dixon and Smirnov (BDS) formula
\cite{Bern:2005iz} and a finite remainder (remainder function), 
which is a function of the cross-ratios of the cusp coordinates for the null polygon.

The amplitudes have been studied intensively from both weak- and strong-coupling sides.
At weak coupling,  recent developments 
using the mixed motive theory have made it possible to evaluate the remainder function for the
six-point amplitudes  up to four-loop level
\cite{Dixon:2014voa}. 
Moreover a method based on the OPE and integrability
has been proposed
to calculate the scattering amplitudes in this theory, which is expected to be applicable to the
intermediate coupling region\cite{Alday:2010ku,Basso:2013vsa,Basso:2013aha,Basso:2014koa}.   

At strong coupling, 
the AdS/CFT correspondence  asserts explicit relations 
between ${\cal N}=4$ super Yang-Mills  theory 
and the superstring theory in AdS$_{5} \times$S$^{5}$.
Alday and Maldacena have thereby proposed
that the MHV amplitude
can be evaluated by the area of the minimal surfaces in AdS 
with  a null-polygonal boundary along the Wilson loop \cite{Alday:2007hr}.
It turns out later that
the remainder function for the null-polygonal  minimal surfaces is
calculated with the help of integrability \cite{Alday:2009yn}. 
Namely it is obtained
by solving the Y-system or the thermodynamic Bethe ansatz (TBA) system
related to certain two-dimensional quantum integrable
systems \cite{Alday:2009dv,Alday:2010vh,Hatsuda:2010cc}.
The cross-ratios are given by the Y-functions at special values of the spectral parameter 
and the remainder function is expressed by the free energy of the TBA system
and the Y-functions.

For the null-polygonal minimal surfaces in AdS$_3$ and AdS$_4$ space-time, the relevant 
integrable systems  are the homogeneous sine-Gordon models \cite{FernandezPousa:1996hi}
with purely imaginary resonance parameters \cite{Hatsuda:2010cc}, 
which are the perturbed  SU($N$)$_k$/U(1)$^{N-1}$ coset conformal field theory (CFT)
 at level $k=2$ and $4$, respectively. 
Around the limit where the null boundary becomes regular polygonal,
corresponding to the UV limit of the two-dimensional systems, 
the remainder functions are calculated analytically for lower point amplitudes 
\cite{Hatsuda:2011ke,Hatsuda:2011jn,Hatsuda:2012pb}.
There, the free-energy part is evaluated by the standard bulk conformal
perturbation theory (CPT). In order to evaluate the Y-functions,  
the $g$-function or the boundary entropy is utilized because
the Y-function itself is not well  incorporated  in quantum field theory.
The boundary CPT then efficiently yields  the analytic expansions of the Y-functions.
 The resultant remainder functions  are observed to be close to the  two-loop results 
 after an appropriate
normalization/rescaling. Numerically, one  also finds that this similarity extends beyond
the UV limit.
The minimal surfaces in these cases, however, give 
the amplitudes with some specific 
kinematic configurations of gluon momenta.

The null-polygonal minimal surface in AdS$_5$ with six cusps is the simplest non-trivial 
example that allows the most general kinematic configuration.
 At strong coupling, the relevant two-dimensional system is  the $\mathbb{Z}_4$-symmetric 
integrable model \cite{Koberle:1979sg, Tsvelick88, Fateev:1990bf} with a boundary 
twist \cite{KBP}.
The remainder function around the UV limit in this case has been studied 
in detail in \cite{Hatsuda:2010vr}.
Although the free-energy part  is analytically evaluated by the bulk CPT 
for the twisted $\mathbb{Z}_4$-parafermion, the expansion of the Y-functions there is
determined by numerical fitting. 
The difference from the  the AdS$_3$  and AdS$_4$ cases 
come from the fact that 
the TBA equations in the AdS$_5$ case have a twist parameter, and it is unclear how
to construct the $g$-function with this twist parameter.

Given the analytic results at weak coupling as well as the OPE method for finite coupling,
the analytic data at strong coupling would provide pieces of
the whole picture of the scattering amplitudes. They would also be useful for a 
check of the finite-coupling analysis. 
In this report, 
we thus decide to 
devote ourselves to the analytic expansions of
the remainder function for the general kinematics. 

In order to overcome the problem mentioned above,  we take below 
another route for the UV expansion of the Y-functions, which does not rely 
on the $g$-function. Instead, our analysis  is based on 
a seminal work by 
Bazhanov, Lukyanov and Zamolodchikov \cite{Bazhanov:1994ft,Bazhanov:1996dr}, 
where  the role  of  quantum monodromy matrix  
is clarified in  the minimal 
CFT, ${\cal M}_{2,2n+3}$, 
perturbed by the $\Phi_{1,3}$ operator.
Most remarkably, a new object in field theories, Baxter's Q operator, is introduced in this work.
They noted  the importance of  the fundamental relations among  the T- and Q-functions, 
 the quantum Wronskian relation \cite{Bazhanov:1996dr}.
The  T- and Y-systems can be regarded as colloraries of this.
The Q operators and the quantum Wronskian relation have also played  
important roles in  the non-equilibrium 
current problem \cite{BLZ:1999},  in the ODE/IM correspondence \cite{Dorey:2007zx},
in the spectral problem of the
AdS$_5$/CFT$_4$ correspondence \cite{Gromov:2010km} and so on.

In this report, we  provide  a yet another application: the  quantum Wronskian relation
is  very efficient in obtaining the analytic expansion of the
Y-functions particularly in the CFT limit.
Specifically,  we apply it, for the first time,  to the $\mathbb{Z}_{4}$-symmetric integrable model
or its lattice regularization.  
The lattice regularization adopted here allows one to  elucidate the analyticity 
of the T-/Y-functions numerically.
The expansion of the Y-functions  around the UV limit is then determined analytically up 
to and including the terms of order (mass)$^{\frac{4}{3}}$.
A series of the higher-order coefficients is also determined recursively.

Combined with the free-energy part, 
the UV expansion of the Y-functions  gives the analytic expansion 
of the  six-point reminder function
for the general kinematic configuration. We also compare the strong-coupling results
with the perturbative ones, and find that the rescaled remainder functions are close to each 
other for large ranges of the parameters. This is in accord with the previous
observations in the AdS$_{3}$ and AdS$_{4}$ cases 
\cite{Brandhuber:2009da,Hatsuda:2011ke,Hatsuda:2011jn,Hatsuda:2012pb}
as well as in the perturbative cases  \cite{Dixon:2013eka,Dixon:2014voa}.

This paper is organized as follows:
In Section 2, we review the TBA-system for the six-point gluon scattering amplitudes 
at strong coupling and express
the remainder functions using Y-functions.
In Section 3, we reconsider  the Y-/T-functions based on the lattice model. 
Taking the scaling limit, we
 derive the TBA system for the  amplitudes.
 In Section 4,  we study the CFT limit of the TBA system. 
 Based on the T-Q relation and the quantum Wronskian relation, 
 we calculate the analytic expansion of the
 Y-function in the CFT limit. The detailed analysis of the asymptotics 
 of the related spectral determinant based on non-linear integral equations
is  studied in Appendix A.  
In Section 5, we apply the analytic expansion of the Y-functions 
to determine the leading expansion
of the remainder function for the six-point amplitudes and compare it 
with the perturbative calculations.

%%%%%%%%%%%%%%%%%%%%%%%%%%%%%%%%%%%%%%%%%%%%%%%%%%%%%%%%%%%%%%%%%%%%%%%%
\section{Y-system and TBA for scattering amplitudes at strong coupling}
%%%%%%%%%%%%%%%%%%%%%%%%%%%%%%%%%%%%%%%%%%%%%%%
%
 Let us  begin  with  a review on the evaluation of the six-point MHV amplitudes 
 at strong coupling using TBA of the twisted $\bbZ_4$-symmetric  integrable model.
%
%---------------------------------------------------------------------------------------------------
\subsection{Hitchin system and Stokes data}
%---------------------------------------------------------------------------------------------------
%
Alday and Maldacena proposed a method of computing
the gluon scattering amplitudes in ${\cal N}=4$ super Yang--Mills
using the AdS/CFT correspondence \cite{Alday:2007hr, Alday:2009yn}.
Consider the   the scalar part of the $n$-point gluon  MHV scattering amplitudes
in the strong coupling limit,  
and factor out 
the contribution of the tree amplitudes.
According to 
\cite{Alday:2007hr}, 
the result  can be evaluated
by computing the area  $A$ of the corresponding classical open string solutions 
in AdS$_5$ spacetime:%
\footnote{
Recently, it was shown that there is another contribution from the S${}^5$ part
of  AdS$_{5} \times$ S$^5$
\cite{Basso:2014jfa}  in addition to the area. However, this contribution is independent 
of the cross-ratios and hence
does not affect the discussions below.
}
\begin{equation*}
\frac{{\text {(amplitude)}}}{\text{(tree)}} 
\sim {\rm e}^{ -\frac{\sqrt{\lambda}}{2\pi} A}  
\end{equation*}
where $\lambda$ denotes the  't Hooft coupling.
A string solution represents a minimal surface whose boundary
is a polygon located on the boundary of AdS$_5$.
The polygon consists of $n$ null edges
given by the $n$ momenta of incoming gluons.
Note that the amplitudes are defined in space-time
with signature (3,1), (2,2) or (1,3). 

The equations of motion of the string under the Virasoro constraints  
are rephrased as the SU(4) Hitchin equations
 for the connections $(A_z,A_{\bar{z}})$ and the adjoint scalar fields 
 $(\Phi_z,\Phi_{\bar{z}})$ with the $\bbZ_4$ automorphism\cite{Alday:2009dv}.
They are equivalently presented by the linear equations for a four component vector 
$q(z,\bz;\zeta)$,
\eqb\label{auxlineq}
\Bigl( D_z+\zeta^{-1}\Phi_z \Bigr) q(z,\bz;\zeta)=0,\qquad
\Bigl( D_\bz+\zeta\Phi_\bz \Bigr)     q(z,\bz;\zeta)=0,
\eqe
with appropriate boundary conditions.  
Here $D_z$ and $D_{\bar{z}}$ denote the covariant derivatives and $\zeta$  
stands for the spectral parameter.
The explicit forms of $D_z$ and $D_{\bar{z}}$ are given in  \cite{Alday:2009dv}. 
The information of  the null polygon 
is encoded in the asymptotic behavior 
of $\Phi_z$, which is diagonalized at infinity 
by an appropriate gauge transformation:  
\eqb
h^{-1}\Phi_z h\Eqn{\to}\frac{1}{\sqrt{2}}
{\text {diag}} \bigl(
P(z)^{1/4},
-iP(z)^{1/4},
-P(z)^{1/4},
iP(z)^{1/4} \bigr ).
\eqe
The polynomial degree of $P(z)$ is $n-4$ and its coefficients 
parameterize the shape of  the polygon.
When $n >4$,  the linear equation necessarily possesses irregular singularity at infinity,
which implies the Stokes phenomena.
 Customarily,  the whole complex plane is divided into sectors,
\eqb
W_k: \frac{\pi(2k-3)}{n}+\frac{4}{n}\arg\zeta <
\arg z < \frac{\pi(2k-1)}{n}+\frac{4}{n}\arg\zeta.
\eqe
We denote by  $s_k(z,\bz;\zeta)$,  the most recessive solution as 
$|z|\to\infty$ in $W_k$.  
One can consistently choose  $(s_k,s_{k+1},s_{k+2},s_{k+3})$ as
a linearly independent basis in $W_k$.
This implies a linear dependent relation  among five neighboring $s_j$'s,
\eqb\label{s_linear}
s_k +s_{k+4}= a_{k} s_{k+1}+  b_{k+1} s_{k+2}+ c_{k+3} s_{k+3} ,
\eqe
 where $b_{k+1}$ and $c_{k+1}$ are some constants. 
Note the periodicity  
 $b_{i+3}= b_i$.
 
The normalization of $s_j(z,\bz;\zeta)$ is fixed such that 
\eqb\label{s_normcond}
\langle s_j,s_{j+1},s_{j+2},s_{j+3}\rangle=1,
\eqe
where 
$\langle s_i,s_j,s_k,s_l\rangle \equiv\det(s_i s_j s_k s_l)$.
The $\bbZ_4$ automorphism results in the following relations for the Stokes data,
\eqb
\label{StokesZ4sym1}
\langle s_k,s_{k+1},s_j,s_{j+1}\rangle(\zeta)
\Eqn{=}\langle s_{k-1},s_k,s_{j-1},s_j\rangle(i\zeta),\\
\label{StokesZ4sym2}
\langle s_j,s_k,s_{k+1},s_{k+2}\rangle(\zeta)
\Eqn{=}\langle s_j,s_{j-1},s_{j-2},s_k\rangle(i\zeta).
\eqe
Below, we confine our argument to  the $n=6$ case where 
\eqb\label{Pofz}
P(z)=z^2-U.
\eqe 
For $W_{j+6}=W_j$, we shall  impose the boundary condition,
\eqb\label{s_identity}
s_{j+6}=\mu^{(-1)^j} s_j .
\eqe
The multiplier $\mu$ is parametrized as
\eqb \label{define_mu}
\mu=e^{i \frac{3}{2} \phi}  , 
\eqe
where $\phi$  is real for solutions in the $(1,3)$  or  in the $(3,1)$ signature
of the four-dimensional space-time
whereas it is purely imaginary for the $(2,2)$ signature.
It also appears, e.g., in the relation among $b_i$,
\eqb\label{b_constraint}
b_1b_2b_3=b_1+b_2+b_3+\mu+\mu^{-1}.
\eqe
%
%-------------------------------------------------------------------------------------
\subsection{Y-functions, TBA and evaluation of area}
%-------------------------------------------------------------------------------------
%
The key ingredients in the following discussion
are the Y-functions, which are
defined explicitly by 
\footnote{
 The Y-functions here are identified with those
in \cite{Alday:2010vh,Hatsuda:2012pb} as
$
Y_1(\theta)=\mu^{-1}[Y_{1,1}^{\rm AMSV}(e^\theta)]^{-1}$, \ $
Y_2(\theta)=[Y_{2,1}^{\rm AMSV}(e^\theta)]^{-1}$, \ $ 
Y_3(\theta)=\mu [Y_{3,1}^{\rm AMSV}(e^\theta)]^{-1}. $
The cusp coordinates which appear below are also related as $x_{a+2} = x_{a}^{\rm AMSV}$. 
}
\eqb
\label{defY1}
Y_1(\theta)\Eqn{=}
-\langle s_2,s_3,s_5,s_6\rangle(e^{\theta}),\\
\label{defY2}
Y_2(\theta)\Eqn{=}
\langle s_1,s_2,s_3,s_5\rangle
\langle s_2,s_4,s_5,s_6\rangle(e^{\theta+\pi i/4}),\\
\label{defY3}
Y_3(\theta)\Eqn{=}Y_1(\theta).
\eqe
Then it was shown in  
\cite{Alday:2009dv} 
that the Hirota bilinear identities (or Pl\"ucker relations),  
as well as the relations  (\ref{s_identity}), 
(\ref{StokesZ4sym1}) and (\ref{StokesZ4sym2}),  lead to the following
Y-system,
\eqb
\label{Ysystem1}
Y_1\left(\theta+\frac{\pi i}{4}\right)
Y_1\left(\theta-\frac{\pi i}{4}\right)
\Eqn{=}1+Y_2(\theta),\\
\label{Ysystem2}
Y_2\left(\theta+\frac{\pi i}{4}\right)
Y_2\left(\theta-\frac{\pi i}{4}\right)
\Eqn{=}\bigl(1+\mu Y_1(\theta)\bigr)
 \bigl(1+\mu^{-1} Y_1(\theta)\bigr).
\eqe
The asymptotic behavior of the Y-functions is
shown to be
\eqb\label{Yasympt}
\log Y_1(\theta)\to |Z|e^{\pm(\theta-i\varphi)},\qquad
\log Y_2(\theta)\to
\sqrt{2}|Z|e^{\pm(\theta-i\varphi)}\nn\\[1ex]
\mbox{for}\quad{\rm Re\,}\theta\to\pm\infty,
\qquad
\varphi-\frac{\pi}{4}<{\rm Im\,}\theta<\varphi+\frac{\pi}{4},
\eqe
where $Z$ is a complex parameter with phase $\varphi$. This 
is related to the moduli parameter $U$ in (\ref{Pofz}) as
\eqb\label{ZUrel}
Z\equiv|Z|e^{i\varphi}
=U^\frac{3}{4}\int_{-1}^1(1-t^2)^\frac{1}{4}dt
=\frac{\sqrt{\pi}\Gamma(\frac{1}{4})}{3\Gamma(\frac{3}{4})}
 U^\frac{3}{4}.
\eqe

We further introduce 
\eqb\label{epsilonandY}
\epsilon(\theta)=\log Y_1(\theta+i\varphi),\qquad
\tepsilon(\theta)=\log Y_2(\theta+i\varphi).
\eqe
The asymptotic behavior (\ref{Yasympt}), 
together with  the assumption of the analyticity of $\epsilon, \tepsilon$ 
 in 
the strip  $\Im \theta \in ( -\frac{\pi}{4},  \frac{\pi}{4})$,  leads
to the integral equations, 
\eqb
\label{TBA1}
\epsilon\Eqn{=}2|Z|\cosh\theta
 +\cK_2\ast\log\bigl(1+e^{-\tepsilon}\bigr)
 +\cK_1\ast\log\bigl(1+\mu e^{-\epsilon}\bigr)
             \bigl(1+\mu^{-1}e^{-\epsilon}\bigr),\\
\label{TBA2}
\tepsilon\Eqn{=}2\sqrt{2}|Z|\cosh\theta
 +2\cK_1\ast\log\bigl(1+e^{-\tepsilon}\bigr)
 +\cK_2\ast\log\bigl(1+\mu e^{-\epsilon}\bigr)
             \bigl(1+\mu^{-1}e^{-\epsilon}\bigr),\qquad
\eqe
where
\eqb\label{kernels}
\cK_1(\theta)=\frac{1}{2\pi\cosh\theta},\qquad
\cK_2(\theta)=\frac{\sqrt{2}\cosh\theta}{\pi\cosh 2\theta},\qquad
\eqe
and the symbol $*$ denotes the convolution, 
$f\ast g=\int_{-\infty}^\infty
d\theta' f(\theta-\theta')g(\theta')$.
These turn out be identical to the TBA 
equations for the $\mathbb{Z}_4$-symmetric  integrable model
\cite{Koberle:1979sg, Tsvelick88, Fateev:1990bf}   twisted by $\mu$.
The equations  (\ref{TBA1}) and (\ref{TBA2}) determine $\epsilon$ and $\tepsilon$ 
in the strip completely.

In the original setting,  the geometric data such as cross-ratios are given first, and then
the area of  the surfaces  should be evaluated.  Below, we slightly deform this logic: 
 the TBA equations are  given first,   then  the cross-ratios  and the area are evaluated second.
Once the TBA equations are solved, 
the coefficient $b_k$  in  eq.~(\ref{s_linear}) and 
the cross-ratios of gluon momenta are given by
\eqb\label{Yspecial}
b_k=Y_1\left(\frac{(k-1)\pi i}{2}\right),\qquad
U_k=1+Y_2\left(\frac{(2k+1)\pi i}{4}\right),
\eqe
for $k=1,2,3$  (\text{mod} 3),
where
\eqb\label{crossratios}
U_1=b_2b_3=\frac{x_{14}^2x_{36}^2}{x_{13}^2x_{46}^2},\qquad
U_2=b_3b_1=\frac{x_{25}^2x_{14}^2}{x_{24}^2x_{15}^2},\qquad
U_3=b_1b_2=\frac{x_{36}^2x_{25}^2}{x_{35}^2x_{26}^2}.
\eqe
The cusp coordinates $x_j$ are related to  the external momenta through
$p_j = x_{j} -x_{j+1}$.
 In the literature, $u_k := 1/U_k$ 
 are often used as a basis of independent cross-ratios for 
the six-point case. 
 The  number of  the independent cross-ratios matches 
that of the parameters in the TBA system $(|Z|, \varphi, \mu)$.

Naively, the values of $Y_j$ outside the analytic strip are necessary
in order to evaluate $U_k \,(1\le k \le 3)$.  
Although this can be accomplished by the analytic continuation in principle,
we can avoid this by a clever choice of quantities. 
For example,  suppose $\varphi$ is negative and small. Two quantities, $b_1$ 
and $U_2=U_{-1}$, are readily calculated by 
(\ref{Yspecial}). Then one evaluates $b_3$ 
by the second equation in  (\ref{crossratios}).
The final piece, $b_2$, is  obtained from (\ref{b_constraint}). 
Given $b_i$, other cross ratios are now accessible via (\ref{crossratios}).
Alternatively, one may also use the Y-system (\ref{Ysystem1}), (\ref{Ysystem2}) 
as recurrence relations to
generate $Y_j(k\pi i/4)$ for any $k \in \bbZ$ from a set of $Y_j(k'\pi i/4)$ in the analytic
strip. 
 As discussed shortly, the Y-functions
have the periodicity $Y_j(\theta + 3\pi i/2) = Y_j(\theta)$, and thus the procedure terminates
after a few steps.

 We are now in position to write down the area $A$ 
of the 6-cusp solutions or 
the scalar magnitude of the gluon scattering amplitudes in the  strong coupling limit.
 Instead of $A$ itself,  we deal with the finite remainder defined by
\eqb
   R = A_{\mbox{\scriptsize BDS}} - A ,
\eqe 
where $A_{\mbox{\scriptsize BDS}}$ is
the all-order ansatz for the MHV  amplitude
proposed by Bern, Dixon and Smirnov\cite{Bern:2005iz},  including the divergent part.
The present formulation then yields
\eqb\label{RemainderFn}
R
\Eqn{=} 
\Delta A_{\rm BDS}
-A_{\mbox{\scriptsize periods}}
-A_{\mbox{\scriptsize free}} , 
\eqe
and each part reads
\eqb\label{R1inU}
\Delta A_{\rm BDS} 
\Eqn{=}-\frac{1}{4}\sum_{k=1}^3 \dilog\left(1 - U_k\right),\\
A_{\mbox{\scriptsize periods}}\Eqn{=}|Z|^2,\\
\label{eq:F}
A_{\mbox{\scriptsize free}} 
\Eqn{=}\frac{1}{2\pi}\int_{-\infty}^\infty d\theta
 \biggl(
 2|Z|\cosh\theta\log\bigl(1+\mu e^{-\epsilon(\theta)}\bigr)
 \bigl(1+\mu^{-1}e^{-\epsilon(\theta)}\bigr)\nn\\
\Eqn{}\hspace{5em}
 {}+2\sqrt{2}|Z|\cosh\theta\log\bigl(1+e^{-\tepsilon(\theta)}\bigr)
 \biggr).
\eqe
The minus of the last term $F = -A_{\rm free}$  coincides with  the free energy 
whereas   $2 |Z|$ is identified with the mass/scale parameter 
 of the $\bbZ_4$-symmetric integrable model.
The overall coupling dependence $\sqrt{\lambda}$ has been omitted above.

Within the framework described above, 
the numerical solutions of  (\ref{TBA1}) and (\ref{TBA2})  yield
explicit evaluation of the  gluon scattering amplitudes\cite{Hatsuda:2010vr} .
 Although the analytic solution to the TBA equations  for generic $Z$ and $\mu$ is 
 beyond our reach,
two limiting cases are accessible  \cite{Zamolodchikov:1989cf}. 
One is the limit where $|Z| \to \infty$.
In this case, the integrable model reduces to a free massive theory, and 
the  free-energy part $A_{\rm free}$ and the Y-functions $Y_j(\theta)$ 
are expanded by multiple integrals. 
Via analytic continuation, this limit is also relevant for 
the amplitudes in the Regge limit \cite{Bartels:2010ej,Bartels:2013dja}.
Another limit is $|Z| \to 0$, which we are interested in here.

When $|Z|$ is strictly zero, 
 $A_{\rm free}$ and $Y_j$ are obtained as the central charge
of the $\bbZ_4$-parafermion theory and a solution to the 
constant Y-system, respectively. Moreover,  for small $|Z|$ the free-energy part is
expanded by the bulk conformal perturbation theory (CPT). 
The Y-functions are expanded by the boundary CPT through the relation to the $g$-function
for $\mu =1$ \cite{Hatsuda:2012pb}, corresponding to the minimal surfaces in AdS${}_4$.
For generic $\mu$, however, it is still unclear how to incorporate $\mu$ in the framework
of the boundary CPT.

 In the following, we take an approach to the problem, which is different from any of the above,
and is 
based on the integrable field theoretical structure proposed 
in \cite{Bazhanov:1996dr}.
This allows us to analytically evaluate the Y-functions for small $|Z|$, 
as shown in section \ref{section:ExpTY}.

%
%-------------------------------------------------------------------------------------
\section{T-functions from the lattice regularization and their scaling limit}
%-------------------------------------------------------------------------------------
%

In this section
we embed the Y-system into another tractable object in integrable systems,
the T-system. 
This  enable us to apply the machinery of the latter 
 to evaluate the Y-functions for small $|Z|$ in the next section.

There are several ways to introduce the T-system which is equivalent to  the Y-system in
(\ref{Ysystem1}) and (\ref{Ysystem2}). 
Here we start  with the lattice regularization\cite{DdV:1987}.
One advantage of this choice is that analyticity assumptions,
necessary to derive the TBA equations, can be checked numerically. 

As is well known, the $\mathbb{Z}_4$ parafermion model is related to
the spin-${1\over2}$ XXZ model.
Its Hamiltonian and spectrum can be studied from the transfer matrix.
In order to define the latter,  
we introduce   $R(v)$, 
the $U_{q}(\widehat{\mathfrak{sl}_2})$  $R$ matrix   of spin $\frac{1}{2}$ 
representation:
\begin{align*} 
R(v) &= 
\begin{pmatrix}
 a(v) &               &                                           &                & \\
           &    b(v) &            c(v)&             &\\
&    c^{-1}(v)&      b(v)    &       &\\
 &               &                 &                a(v)  &
\end{pmatrix} ,\\
a(v)&=\frac{\sin(v+ \gamma) }{  \sin \gamma   } , \qquad
b(v)=\frac{\sin(v)}{  \sin \gamma   } ,\qquad
c(v)={\rm e}^{-v} ,
\end{align*}
where  $ q={\rm e}^{i\gamma}$.
Let   $V^{(m)}$ be  the $m+1$ dimensional $U_{q}(\mathfrak{sl}_2)$ module
and
$V^{(m)}(v)$ 
be  the corresponding $U_{q}(\widehat{\mathfrak{sl}_2})$  module.   
By  $V^{(m)}_i(v)$ we mean its i-th copy.
The $R$ matrix acting on $V_i^{(1)}(v_i)\otimes V^{(1)}_j(v_j)$  is denoted by $R_{i,j}(v_i-v_j)$.

Then  one can construct  an inhomogeneous  transfer matrix ${\mathbf T}_1(x)$ acting on 
$2N$ sites by,
\eqb
{\mathbf T}_1(x) ={\rm Tr}_0   D_{\phi} R_{0, 2N}(ix+ i\Lambda)  R_{0, 2N-1}(ix -i\Lambda)  
\cdots  R_{0, 2}(ix+ i\Lambda) R_{0, 1}(ix- i\Lambda) ,
\eqe
where suffix 0 denotes the auxiliary space.
The spectral  parameter $x$  is set via $v=ix$ for later convenience.
We have also  introduced the diagonal twist matrix 
$D_{\phi}=[{\rm e}^{-\frac{ \phi}{2}i } , {\rm e}^{\frac{ \phi}{2}i } ]$
under the trace. 
The quantity $\mu={\rm e}^{i\frac{3}{2}\phi}$ is to be identified with the multiplier in 
(\ref{define_mu}).

To diagonalize  ${\mathbf T}_1$, Baxter\cite{Baxter:1972} ingeniously introduced
an operator ${\mathbf Q}$  which commutes with ${\mathbf T}_1(x)$.
 They satisfy  Baxter's TQ relation,
\eqb\label{BaxterTQ}
{\mathbf T}_1(x) {\mathbf Q}(x) = \Phi(x+ i\frac{\gamma}{2})  {\mathbf Q}(x-i\gamma)+
\Phi(x- i\frac{\gamma}{2}) {\mathbf Q}(x+i\gamma) ,
\eqe
where 
\begin{equation*}
\Phi(x) = \bigl( -4 \sinh (x-\Lambda) \sinh(x+\Lambda) \bigr)^N.
\end{equation*}
Below we shall consider  ${\mathbf T}_1$ and  ${\mathbf Q}$  on their common eigenspace, thus
we do not distinguish operators from their eigenvalues.
The eigenvalue of  ${\mathbf Q}$  is explicitly written with  a set of Bethe roots  
$\{x_j(\phi)\}$ with twist $\phi$,
\eqb
\qquad {\mathbf Q}(x) ={\rm e}^{\frac{x}{2\gamma}\phi }\prod_{j=1}^m 
2\sinh(x-x_j(\phi)).  \nonumber 
\eqe
This, together with (\ref{BaxterTQ}), parameterizes the eigenvalue of the transfer matrix.

The fusion method generates a series of vertex models such that the auxiliary space 
is $V^{(j)}(v)$.
Let ${\mathbf T}_j(x)$ be the corresponding  inhomogeneous transfer matrix, 
with a suitable normalization.%
\footnote{Note suffix $j$ is twice of that in \cite{Bazhanov:1996dr}.} 
By construction $\{ {\mathbf T}_j(x) \}$ constitute a commutative family and 
they satisfy the T-system of SU(2) type, 
\begin{equation}
{\mathbf T}_j(x+ \frac{\gamma}{2}i) {\mathbf T}_j(x- \frac{\gamma}{2} i) 
=f_j(x)+{\mathbf T}_{j+1}(x) 
{\mathbf T}_{j-1} (x), 
\qquad j \in \mathbb{N}
\label{tsys0}
\end{equation}
where 
\begin{align*}
{\mathbf T}_0(x) &:=  \Phi(x)  , \qquad 
f_j(x)={\mathbf T}_0(x+ \frac{j+1}{2} \gamma i) {\mathbf T}_0(x-\frac{j+1}{2} \gamma i).
\end{align*}
Note the periodicity of ${\mathbf T}_j(x)$ in the present normalization is
\begin{equation}
{\mathbf T}_j(x+\pi i) = {\mathbf T}_j(x).
\label{eq:periodicity-T}
\end{equation}

There is an additional relation when $q$ is at a root of unity, which is crucial in
obtaining a closed set of functional relations.
From now on we fix 
\begin{equation}
\gamma = \frac{2\pi}{3}.
\label{eq:gamma} 
\end{equation}
Then the desired relation is
\begin{equation}
{\mathbf T}_{3}(x) ={\mathbf T}_{1}(x)+
{\mathbf T}_0(x) \bigl(\mu+\mu^{-1}).
\label{linearrel}
\end{equation}
The equation for $j=2$ in (\ref{tsys0}) can be thus  rewritten as
\begin{equation}\label{T2ndeq}
{\mathbf T}_2(x+ \frac{\pi}{3}i) {\mathbf T}_2(x- \frac{\pi}{3} i) =
({\mathbf T}_{1}(x)+\mu {\mathbf T}_{0}(x) ) 
({\mathbf T}_{1}(x)+\mu^{-1} {\mathbf T}_{0}(x) ),
\end{equation}
thereby yielding a closed functional relations among   ${\mathbf T}_1$ and   ${\mathbf T}_2$.

Below we will  show  that  (\ref{tsys0}) for $j=1$ and (\ref{T2ndeq}) 
can be transformed into TBA equations.
Before doing this,  we elucidate the analytic properties of  ${\mathbf T}_j$
deduced from numerics,  as a merit in  the lattice regularization.
By definition,  $ {\mathbf T}_j$ has $2N$ zeros and  has no poles  in complex $x$ plane.
Led by numerical observations we conjecture that 
all zeros of $ {\mathbf T}_1(x)$ are on the $\Im x=\frac{\pi}{2}$ 
line while those of $ {\mathbf T}_2(x)$ are on the real axis in the ground state.
Below, quantities which have no zeros and poles in the strip including the real axis will play an
important role.
We thus define 

\begin{equation}
{\mathbf T}^{\vee}_j(x) =
{\mathbf T}_j(x+\frac{(j-1)\pi}{2}i)  \qquad j=1,2.
\label{defshiftTj}
\end{equation}
Then the closed functional relations now read 
\begin{align*}
{\mathbf T}^{\vee}_1(x+ \frac{\pi}{6}i) {\mathbf T}^{\vee}_1(x-  \frac{\pi}{6} i) 
&=f_1(x+\frac{\pi}{2}i )+ {\mathbf T}_{0}(x+\frac{\pi}{2}i){\mathbf T}^{\vee}_{2}(x) , \\
{\mathbf T}^{\vee}_2(x+ \frac{\pi}{6}i) {\mathbf T}^{\vee}_2(x- \frac{\pi}{6}i) &=
({\mathbf T}^{\vee}_{1}(x)+\mu {\mathbf T}_{0}(x) ) 
({\mathbf T}^{\vee}_{1}(x)+\mu^{-1} {\mathbf T}_{0}(x) ).
\end{align*}
By adopting the change of variables,
\begin{equation*}
{\mathbf  Y}_1(x)  = \frac{{\mathbf T}^{\vee}_1(x)}{{\mathbf T}_0(x)} , \qquad
{\mathbf  Y}_2(x)  = 
\frac{{\mathbf T}_{0}(x+\frac{\pi}{2}i){\mathbf  T}_2^{\vee}(x)}{f_1(x+\frac{\pi}{2}i)}, 
\end{equation*}
it is  easily checked that ${\mathbf  Y}_1$ and ${\mathbf  Y}_2$ satisfy the same Y-system 
as  (\ref{Ysystem1}) and (\ref{Ysystem2})
if  $\theta=\frac{3 x}{2}$.

Thanks to the knowledge on the zeros of the T-functions, 
one concludes that 
 ${\mathbf  Y}_1(x)$ possesses poles of order $N$ at $x=\pm \Lambda$ while
${\mathbf  Y}_2(x)$ possesses poles of order $N$ at $x=\pm \Lambda \pm \frac{\pi}{6}i$.
There are no other poles or zeros of  ${\mathbf  Y}_1(x)$, ${\mathbf  Y}_2(x)$  in the strip
$\Im x \in [-\frac{\pi}{6}, \frac{\pi}{6}]$.   This motivates us to define the pole-free  functions,
\begin{equation*}
{\widetilde{\mathbf  Y}}_1(x)  = {\mathbf  Y}_1(x) \mathbf{D}_1 (x) , \qquad
{\widetilde{\mathbf  Y}}_2(x)  ={\mathbf  Y}_2(x)  \mathbf{D}_2(x) ,
\end{equation*}
where  
\begin{equation*}
\mathbf{D}_1 (x) =\bigl(  - \tanh  \frac{3}{4}(x-\Lambda) 
 \tanh  \frac{3}{4}(x+\Lambda) \bigr)^N , \qquad
\mathbf{D}_2 (x)=   \mathbf{D}_1(x+i\frac{\pi}{6})  \mathbf{D}_1 (x-i\frac{\pi}{6}) .
\end{equation*}

One then derives the functional equations, 
\begin{align*}
\frac{{\widetilde{\mathbf  Y}}_1(x+\frac{\pi}{6}i)  
  {\widetilde{\mathbf  Y}}_1(x-\frac{\pi}{6}i)}{ {\widetilde{\mathbf  Y}}_2(x)  }
&=\Bigl(1+ \bigl({\mathbf Y}_2(x) \bigr)^{-1} \Bigr) , \\
\frac{{\widetilde{\mathbf  Y}}_2(x+\frac{\pi}{6}i)   
 {\widetilde{\mathbf  Y}}_2(x-\frac{\pi}{6}i)}{ \bigl({\widetilde{\mathbf  Y}}_1(x)\bigr)^2  }
&=(1+ \mu {\mathbf Y}_1^{-1}(x))(1+ \mu^{-1} {\mathbf Y}_1^{-1}(x)),
\end{align*}
where  both sides do not have any zeros or poles in the strip.
Thanks to the analyticity, one  arrives at   
\begin{align*}
\log {\mathbf  Y}_1(x) &= -\log  \mathbf{D}_1 (x)  +
K_1*\log \Bigl(1+\frac{\mu^{-1}}{{\mathbf  Y}_1} \Bigr)\Bigl(1+\frac{\mu}{{\mathbf Y}_1} \Bigr)(x)
+ K_2*\log\Bigl(1+  \frac{1}{{\mathbf  Y}_2} \Bigr) (x) , \\
\log {\mathbf  Y}_2 (x)&=-\log  \mathbf{D}_2 (x)  +
K_2*\log \Bigl(1+\frac{\mu^{-1}}{{\mathbf  Y}_1} \Bigr)\Bigl(1+\frac{\mu}{{\mathbf Y}_1} \Bigr)(x)
+ 2K_1*\log\Bigl(1+  \frac{1}{{\mathbf  Y}_2} \Bigr)  (x) ,
\end{align*}
where
\begin{equation*}
K_1(x)=\frac{3}{4 \pi \cosh\frac{3}{2}x} ,
\qquad
K_2(x)=\frac{3 \cosh \frac{3}{2} x}{\sqrt{2} \pi  \cosh 3 x} .
\end{equation*}

Now consider the following scaling limit, 
\eqb
\lim_{N \rightarrow \infty} 4 N {\rm e}^{-\frac{3}{2}\Lambda} = 2|Z|  =\ell . \label {Lambdalimit}
\eqe
In this limit, the driving terms become
\begin{align}
\lim_{N \to \infty} \log \mathbf{D}_1(x)=-\ell \cosh\(\frac{3x}{2} \),\quad
\lim_{N \to \infty} \log \mathbf{D}_2(x)=-\sqrt{2} \ell \cosh\(\frac{3x}{2} \).
\end{align}
We denote the Y-functions in the scaling limit by $\mathbf{Y}^\text{sc}_j(x)$.
If changing the variables as $\theta=\frac{3}{2}x$, we recover (\ref{TBA1}) and (\ref{TBA2}) 
by the identification,
\begin{equation*}
\log {\mathbf Y}_1^\text{sc}( x) =  \epsilon(\theta) ,
\qquad
\log {\mathbf Y}_2^\text{sc}( x) = \tepsilon(\theta).
\end{equation*}
We also define the T-functions in the scaling limit by
\begin{align}
\mathbf{T}^\text{sc} _j(x)= 
\lim_{N\to \infty} {\rm e}^{-2\Lambda N} \mathbf{T} _j(x) ,
\end{align}
where especially $ \mathbf{T}^\text{sc} _0(x)= 1$.
In the scaling limit,  the relations between the Y-functions and 
the T-functions are drastically simplified,
\eqb
{\mathbf Y}_1^\text{sc}(x)  = {\mathbf T}_1^{\text{sc}}(x) ,  \qquad 
{\mathbf Y}_2^\text{sc}(x)  = {\mathbf T}_2^{\text{sc}}(x+\frac{\pi i}{2}). 
\label{eq:YT-rel-sc}
\eqe 

For later use, we shall also discuss the scaling limit of $\mathbf{Q}$.
The Bethe ansatz roots are roughly classified into two clusters, 
$x_j^{\ell} \sim -\Lambda$  and  $x_j^{r} \sim  \Lambda$.
We thus adopt parameterizations,
\begin{equation*}
\tilde{x}_j^r(\phi)= x_j^r(\phi) -\Lambda ,
\qquad
\tilde{x}_j^\ell (\phi)= x_j^\ell (\phi)+\Lambda ,
\end{equation*}
and  
\begin{equation*}
\lambda = {\rm e}^{x}={\rm e}^{\frac{2}{3}\theta} ,  \qquad
\lambda_j^ r(\phi)  = {\rm e}^{\tilde{x}_j^r(\phi) } (2N)^{\frac{2}{3}},  \qquad
\lambda_j^ {\ell}(\phi) = {\rm e}^{-\tilde{x}_j^\ell (\phi)} (2N)^{\frac{2}{3}}.
\end{equation*}
Then the  scaling limit  of $\mathbf{Q}^{\text{sc}}$ reads 
\begin{align} \label{scaleQ}
\mathbf{Q}^{\text{sc}}(\lambda)&=
\lim_{N \rightarrow \infty, \ell=\text{fixed}}  {\rm e}^{-N\Lambda } \mathbf{Q}(x) \nonumber \\
&= {\mathfrak C}(\phi) \lambda^{\frac{3\phi}{4 \pi}}  
\prod_j \Bigl(1- \bigl(\frac{ (\frac{\ell}{2})^{\frac{2}{3}}\lambda}{\lambda_j^ r(\phi)}\bigr)^2\Bigr)
\prod_j \Bigl(1-  
\bigl(\frac{  (\frac{\ell}{2})^{\frac{2}{3}}\lambda^{-1}}{\lambda_j^ \ell (\phi)}\bigr)^2\Bigr) ,
\end{align}
where we assumed the numbers of roots in the left and the right clusters are  
identically equal to $\frac{N}{2}$.
The prefactor stands for 
\eqb
{\mathfrak C}(\phi)=\prod_j 
 {\rm e}^{\tilde{x}_j^r(\phi)-\tilde{x}_j^\ell (\phi)+\pi i}.
\eqe
This Q-function plays an important role for studying analytical properties of the Y-functions.    
%
%-------------------------------------------------------------------------------------
\section{Expansions of T-/Y-functions around CFT limit}
\label{section:ExpTY}
%-------------------------------------------------------------------------------------
In this section, we consider the expansions of the T- and Y-functions around $\ell=2|Z|=0$.
 As mentioned in the previous section, such expansions for the minimal surface 
in AdS$_3$ or AdS$_4$ were studied in detail 
in \cite{Hatsuda:2011ke, Hatsuda:2011jn, Hatsuda:2012pb} 
 through the bulk and boundary conformal perturbation theory.
Also, in \cite{Hatsuda:2010vr}, the expansions of the Y-functions for the six-point case 
in AdS$_5$ were considered,
but there remained an unknown function  
$Y_j^{(1,0)}(\phi)$ at order $\ell^{4/3}$. 
Our goal here is to determine the analytic form of this unknown function.
The key idea is to use the quantum Wronskian, which is naturally derived 
from the discretized lattice regularization
reviewed in the previous section.
The quantum Wronskian determines not only 
$Y_j^{(1,0)}(\phi)$ but also the expansion of $Y_j$ in the CFT limit defined below.
For the time being, we set $Z$ to be real, i.e., $\varphi=0$.

\subsection{General argument}
Let us start with a general argument on the expansions of the T- and Y-functions.
Solutions to the Y-system have a periodicity as conjectured first in \cite{Zamolodchikov:1991et},
and it played an important role in the analysis of the perturbed  CFT 
(see, e.g. \cite{Zamolodchikov:1989cf}).
Here we have an  apparent periodicity\footnote{%
Note that the periodicity can be proved even without  information
on its lattice origin.
The cluster algebraic structure is shown to be essential 
\cite{Fomin, Keller:2008, IIKKN:2013a, IIKKN:2013b}.
The argument here is meant to explain the periodicity in a simple manner.
}
inherited from the underlying lattice model (see \eqref{eq:periodicity-T}), and it 
motivates the expansions of  $Y_j( \theta)$ and $T_j(\theta)$,
\eqb
Y_j(\theta)  =\sum_{p=0}^{\infty} 2 Y^{(p)}_j \ell^{\frac{4}{3}p} \cosh \frac{4p}{3} \theta ,
\qquad
T_j(\theta)  =\sum_{p=0}^{\infty} 2 T^{(p)}_j  \ell^{\frac{4}{3}p}   \cosh \frac{4p}{3} \theta  ,
\label{eq:YT-exp1}
\eqe
 where $Y_j(\theta)$ ($T_j(\theta)$) is $\mathbf{Y}_j^\text{sc}(x)$ ($\mathbf{T}_j^\text{sc}(x)$)  
 as a function of $\theta$.
The conformal perturbative argument suggests that the coefficients are 
also expanded around $\ell=0$,%
\footnote{
When $\mu = 1$, the Y-functions for the $n$-point amplitudes, $Y_{a,s}(\theta)$ 
$(a=1,2; s=1,..., n-5)$,
have the quasi-periodicity $Y_{a,s}(\theta +n \pi i /4) = Y_{a,n-4-s}(\theta)$. 
They are also expanded \cite{Hatsuda:2012pb} as 
$Y_{a,s}(\theta) = \sum_{p,q} y_{a,s}^{(p,2q)} \ell^{(p+2q)(1-\Delta)} \cosh(4p \theta/n)$ 
with $\Delta = (n-4)/n$.
For $n=6$, the above quasi-periodicity is promoted to the periodicity
$Y_{a}(\theta +n \pi i /4) = Y_{a}(\theta)$, where $Y_a := Y_{a,1}$,  and thus only 
$p$ even is allowed. 
This gives the expansion of the  form as in (\ref{eq:YT-exp1}) and  (\ref{YTpq}).
}

\eqb\label{YTpq}
 Y^{(p)}_j  = \sum_{q} Y^{(p,q)}_j \ell^{\frac{4}{3}q}
\qquad
T^{(p)}_j  = \sum_{q} T^{(p,q)}_j \ell^{\frac{4}{3}q} .
\label{eq:YT-exp2}
\eqe
The central issue here is  to determine these coefficients.

%-------------------------------------------------------------------------------------
\subsection{Quantum Wronskian and  CFT limit }
%-------------------------------------------------------------------------------------
We remind that Baxter's TQ-relation (\ref{BaxterTQ}) 
is a second order difference equation, and there are two linearly  independent solutions.
Let $\{ x_j(-\phi) \}$ be a set of BAE roots for negative twist $-\phi$, then the
second solution reads   
$\bar{{\mathbf Q}}(x)={\rm e}^{-\frac{x}{\gamma}\phi} \prod_j 2 \sinh(x-x_j(-\phi))$.
Its scaling limit is obtained from  (\ref{scaleQ}) by $\phi\rightarrow -\phi$.

There exist remarkable relations (the quantum Wronskian relations)  
between such two independent solutions 
of Baxter's TQ relation and
the fusion transfer matrices\cite{Bazhanov:1996dr},
\begin{align}
2i \sin \frac{\phi}{2} \mathbf{T}_j(x)=&
{\mathbf Q}(x-i\frac{j+1}{3}\pi ) \bar{{\mathbf Q}}(x+i\frac{j+1}{3}\pi) \notag  \\
&-{\mathbf Q}(x+i\frac{j+1}{3}\pi ) \bar{{\mathbf Q}}(x-i\frac{j+1}{3}\pi), 
\qquad j=0,1,2,\cdots.
\label{qWronskian}
\end{align}
The same relation  naturally arises in the context of  the massive generalization of 
the ODE/IM 
correspondence\cite{LZ:2010}. A little bit different view from the lattice model 
is remarked in \cite{Pronko:1999}.

Below we will show that coefficients 
 $Y^{(p,0)}_j\,(j=1,2)$ are determined by applying the above relations.
We start with the scaling limit of (\ref{qWronskian}),
\eqb\label{qWronskianscaling}
2i \sin \frac{\phi}{2} \mathbf{T}^\text{sc} _j(\lambda)=
{\mathbf Q}^\text{sc}(\lambda q^{-\frac{j+1}{2}})
 \bar{{\mathbf Q}}^\text{sc}(\lambda q^{\frac{j+1}{2}})
-{\mathbf Q}^\text{sc}(\lambda q^{\frac{j+1}{2}})
\bar{{\mathbf Q}}^\text{sc}(\lambda q^{-\frac{j+1}{2}}) .
\eqe
Note $q={\rm e}^{\frac{2\pi}{3}i}$.
We further take the CFT limit $\ell \rightarrow 0$ 
with the shift $\lambda \rightarrow\bigl(\frac{ \ell}{2}\bigr)^{-\frac{2}{3}} \lambda$.
Let us define scaled  functions,
\begin{align}
&\mathbf{T}^\text{CFT} _j(\lambda^2) =\lim_{\ell \rightarrow 0}  
\mathbf{T}^\text{sc} _j(\bigl(\frac{ \ell}{2}\bigr)^{-\frac{2}{3}} \lambda) ,\\
&\mathbf{A}(\lambda) = \lim_{\ell \rightarrow 0} 
 \prod_j \Bigl(1- \bigl(\frac{\lambda}{\lambda^r_j(\phi)}\bigr)^2 \Bigr) ,
&\bar{\mathbf{A}}(\lambda) = \lim_{\ell \rightarrow 0} 
 \prod_j \Bigl(1- \bigl(\frac{\lambda}{\lambda^r_j(-\phi)}\bigr)^2 \Bigr).
\end{align}
Then the quantum Wronskian relation takes  the form, 
\eqb\label{qWronskianCFT}
2i \sin \frac{\phi}{2} \mathbf{T}^\text{CFT}_j(\lambda^2)=
{\rm e}^{ i\frac{\phi}{2}(j+1)} {\mathbf A}(\lambda q^{-\frac{j+1}{2}})
\bar{{\mathbf A}}(\lambda q^{\frac{j+1}{2}})
-{\rm e}^{- i\frac{\phi}{2}(j+1)} {\mathbf A}(\lambda q^{\frac{j+1}{2}})  
\bar{{\mathbf A}}(\lambda q^{-\frac{j+1}{2}}), 
\eqe
where we used ${\mathfrak C}(\phi){\mathfrak C}(-\phi)=1$,  
as $\{x_j(\phi) \}=\{ -x _j(-\phi)\}$  resulting  form the Bethe ansatz equations.
The  same limit  of the Y-functions is denoted by 
$Y_{j}^\text{CFT} (\lambda^2)= \lim_{\ell \rightarrow 0}  Y_j(\theta-\log \frac{\ell}{2})$, 
which has an obvious expansion from \eqref{eq:YT-exp1} and \eqref{eq:YT-exp2}, 
\eqb
Y_{j}^\text{CFT}(\lambda^2)=2 Y_{j}^{(0,0)}+ \sum_{p=1}^{\infty}   
 Y_{j}^{(p,0)} 2^{\frac{4p}{3}}\lambda^{2p}.
\eqe

In the literature,  the limit, $\ell \to 0$  without rescaling in  the spectral
parameter is also referred to as the CFT limit.
In this paper, we call it  the UV limit in order to avoid confusion.
In the CFT limit, the mass terms in the TBA equations become chiral,
and we are left with the massless TBA system. 

Our strategy here is to use the quantum Wronskian relation to evaluate 
$Y_{j}^\text{CFT} (\lambda^2)$ and then read off
the coefficients  $Y_{j}^{(p,0)}$.
Let the ``$n$-th  (inverse)  moment"  of BAE roots be $a_n$  and $\bar{a}_n$,
\eqb
a_n =\lim_{\ell \rightarrow 0}  \frac{1}{n} \sum_j 
 \Bigl(  \frac{1}{( \lambda^r_j(\phi))^2} \Bigr)^{n}, \qquad
\bar{a}_n = \lim_{\ell \rightarrow 0}  \frac{1}{n} \sum_j  
\Bigl(  \frac{1}{ (\lambda^r_j(-\phi) )^2}\Bigr)^{n} .
\eqe
Then  the following expansion is valid for $0\le \phi \le 2(\pi-\gamma)$,
\eqb\label{Aexpansion}
\ln {\bf A}(\lambda ) = - \sum_{n=1}^{\infty} a_n \lambda^{2n}, \qquad 
\ln\bar{ {\bf A}}(\lambda ) = - \sum_{n=1}^{\infty} \bar{a}_n \lambda^{2n}.
\eqe
In the CFT limit, the relations between the Y-functions and the T-functions are simple,
\eqb
Y_{1}^\text{CFT} (\lambda^2)  = {\mathbf T}_1^{\text{CFT}}(\lambda^2),  \qquad 
Y_{2}^\text{CFT} (\lambda^2)  = {\mathbf T}_2^{\text{CFT}}(-\lambda^2).
\eqe 
Using the quantum Wronskian (\ref{qWronskianCFT}), one can relate the
coefficients $Y_{j}^{(p,0)}$ to $a_n$.
To do so, let us
introduce polynomials $p_n(y)$ in $\{y_1, y_2, \cdots\}$ 
by
\begin{equation}
\exp(\sum_{n=1}^{\infty}y_n x^n) = \sum_{\nu=0}^{\infty}p_\nu(y)x^\nu .
\end{equation}
Here $p_0(y)=1$, $p_1(y)=y_1$, $p_2(y)={1\over2}y_1^2+y_2$ etc.
Then from (\ref{qWronskianCFT}), one obtains 
\begin{align}
Y_j^{(0,0)}&={\sin\left({j+1\over2}\phi\right) \over 2\sin{\phi\over2}},\\
Y_j^{(n,0)}&=(-1)^{n(j-1)} 2^{-{4\over3}n}
\sum_{\nu=0}^{n} p_\nu(-a)p_{n-\nu}(-\bar{a})
\frac{\sin{j+1\over2} \left({4\pi\over3}(n-2\nu)+\phi\right)}
{\sin{\phi\over2}}, \quad (n\geq 1).
 \nonumber 
\end{align}
Some examples are listed
as follows, 
\begin{align}\label{Y1p0}
Y_{1}^{(0,0)} &=  \cos \left(\frac{\phi }{2}\right) ,
\nn \\
2^{\frac{4}{3}} Y_{1}^{(1,0)} &=
-\frac{1}{\sin \left(\frac{\phi }{2}\right)}\left(a_1 \cos \left(\phi +\frac{\pi
  }{6}\right)-\bar{a}_1 \cos \left(\frac{\pi }{6}-\phi \right)\right), \\
2^{\frac{8}{3}}  Y_{1}^{(2,0)} &= 
\frac{1}{2 \sin \left(\frac{\phi }{2}\right)} \left(2 a_1 \bar{a}_1 \sin
  (\phi )+\left(\bar{a}_1^2-2 \bar{a}_2\right) \cos \left(\phi +\frac{\pi
  }{6}\right)
 -\left(a_1^2-2 a_2\right) \cos \left(\frac{\pi
  }{6}-\phi \right)
  \right), \nn \\
 2^4Y_{1}^{(3,0)} &= 
  \frac{1}{6  \sin \left(\frac{\phi }{2}\right)}  
 \Bigl(\left(-\bar{a}_1^3+6
  \bar{a}_1 \bar{a}_2-a_1^3+6 a_2 a_1-6 (a_3+ \bar{a}_3)\right) \sin (\phi )  \nn \\
  &  -3\left(a_1^2-2 a_2\right) \bar{a}_1 \cos \left(\phi +\frac{\pi }{6}\right)+3
  a_1 \left(\bar{a}_1^2-2 \bar{a}_2\right) \cos \left(\frac{\pi }{6}-\phi
  \right)\Bigr) ,\nn
\end{align}
and 
\begin{align}\label{Y2p0}
 Y_{2}^{(0,0)} &= \cos (\phi )+ \frac{1}{2} ,
 \nn \\
2^{\frac{4}{3}} Y_{2}^{(1,0)} &=\left(\bar{a}_1+a_1\right) (2 \cos (\phi )+1) , \\
2^{\frac{8}{3}}  Y_{2}^{(2,0)} &=
\frac{1}{2} \left(2 a_1 \bar{a}_1+\bar{a}_1^2-2 \bar{a}_2+a_1^2-2 a_2\right) (2
  \cos (\phi )+1) ,\nn \\
2^4Y_{2}^{(3,0)} &=\frac{1}{6} \left(\left(\bar{a}_1+a_1\right)
  \left(\left(\bar{a}_1+a_1\right){}^2-6 \left(\bar{a}_2+a_2\right)\right)+6
  \left(\bar{a}_3+a_3\right)\right) (2 \cos (\phi )+1). \nn
\end{align} 
Thus once the ``moments "$a_n$ and $\bar{a}_n$ are  known,   
$Y_{j}^{(p,0)}$ are easily evaluated.
These ``moments"  satisfy the discrete Wiener-Hopf equations\cite{Dorey:2007zx}.  
They are obtained from  $j=0$ case  
in  (\ref{qWronskianCFT})
by expanding the both sides  order by order in $\lambda^2$.
Explicitly they are of  the form,
\begin{equation}
\sum_{\nu=0}^{n}p_n(-a)p_{n-\nu}(-\bar{a})
\sin{1\over2}\left({4\pi \over3}(n-2\nu)+\phi\right)=0.
\end{equation}
Since $p_n(-a)$ is of the form $-a_n+\cdots$
with the ellipses being 
a polynomial in $a_1,\cdots, a_{n-1}$, the above relation
reduces to 
\eqb\label{discreteWH}
\sin (n \gamma -\frac{\phi}{2}) a_n -
\sin (n \gamma+\frac{\phi}{2}) \bar{a}_n  ={\mathfrak r}_n(\phi),
\eqe
where $\gamma$ is defined in (\ref{eq:gamma}).
The explicit forms of  the first few  ${\mathfrak r}_n(\phi)$  read
\begin{align*}
{\mathfrak r}_1(\phi)&=0 , \\
{\mathfrak r}_2(\phi)&=\frac{\sin\frac{\phi}{2}}{2} \Bigl( a_1^2+ \bar{a}_1^2 
+2 \cos 2\gamma  a_1 \bar{a}_1\Bigr) ,\\
{\mathfrak r}_3(\phi)&=\frac{1}{6} \left(\bar{a}_1^3+a_1^3\right) \sin \left(\frac{\phi
  }{2}\right)-\left(\bar{a}_1 \bar{a}_2+a_1 a_2\right) \sin \left(\frac{\phi
  }{2}\right)    \\
  &+a_2 \bar{a}_1 \cos \left(\frac{1}{6} (\pi -3 \phi )\right)-a_1
  \bar{a}_2 \cos \left(\frac{1}{6} (3 \phi +\pi )\right).
\end{align*}
In general ${\mathfrak r}_n$ is a polynomial of $a_j$ and  $\bar{a}_j$  for   $1\le j \le n-1$.

Remarkably,  they are sufficient to determine  $a_n(\phi)$ \cite{BLZ:1999, Dorey:2007zx}.
To be precise, we need two assumptions to accomplish this.
First, we presume that $a_n$ ($\bar{a}_n)$  is analytic for $\phi >  -\frac{2}{3}\pi $
($ \phi <\frac{2}{3}\pi  $).  

This is consistent with the observation from BAE:
as $\phi \rightarrow -\frac{2}{3}\pi $ one of the roots   $\lambda^r \rightarrow  0$ thus
$a_n$ diverges.
This implies the existence of overlapping of 
the analytic strips near the origin of $\phi$  for both $a_n$ and $\bar{a}_n$.
Second,  the following asymptotic behavior of  $a_n$  as $\phi \rightarrow \infty$ is postulated,
\eqb\label{an:asymp}
a_n  \sim \alpha_n \bigl(\frac{\phi}{2\pi}\bigr)^{1- \frac{4n }{3}}, 
\eqe
where  the coefficient reads  explicitly,
\eqb \label{alphanso1}
\alpha_n  =\frac{ \Gamma(\frac{n}{3}) \Gamma(\frac{2n}{3}-\frac{1}{2})}{4 \pi^{\frac{1}{2}} n!}
\Bigl(
\pi^{\frac{1}{2}}
\frac{\Gamma( \frac{1}{4} )}{ \Gamma(\frac{3}{4})}
\Bigr)^{-\frac{4 n}{3}}.
\eqe
This is deduced from the analysis on  the Bethe ansatz equations in the large 
$\phi $ limit, as shortly
discussed in Appendix  A. 
The first member,  $\alpha_1$ can  also be fixed by the expansion of the Y-functions
for the AdS${}_4$ case corresponding to $\phi =0$ \cite{Hatsuda:2012pb}.

Once these assumptions are taken for granted,  we can successively determine $a_n$.
The first order equation is  simply  solvable and one finds
\eqb\label{a1}
a_1(\phi) =
\frac{\Gamma(\frac{1}{3} + \frac{\phi}{2\pi})}{\Gamma(\frac{2}{3} + \frac{\phi}{2\pi})} \alpha_1.
\eqe
For  $\Re \phi>0$,  the second moment is given explicitly, 
\eqb
a_2(\phi)=\frac{3\alpha_1^2  }{8\pi^3}
\frac{\Gamma(\frac{2}{3}+\frac{\phi}{2\pi})}{\Gamma(\frac{1}{3}+\frac{\phi}{2\pi})} 
\int_{-\infty}^{\infty} 
\frac{dx}{2\pi} \frac{\sinh \frac{x}{2}}{x+i\phi} \bigl( \Gamma(\frac{1}{3}+\frac{ix}{2\pi})
\Gamma(\frac{1}{3}-\frac{ix}{2\pi}) \bigr)^3.    \label{a2sol1} 
\eqe
For $\Re \phi<0$,   $a_2(\phi)$ is evaluated by the analytic continuation of the above expression.
A  double integral formula for the third moment $a_3$ is derived similarly.
For the evaluation up to $Y_j^{(3,0)}$, however,  the explicit forms of $a_3$ and 
$\bar{a}_3$  are dispensable.
This is due to the fact that 
only the sum, $a_3+\bar{a}_3$, appears  in  $Y_j^{(3,0)}$ and this sum also appears in 
the condition (\ref{discreteWH}) for $n=3$,  in the special case $\gamma=\frac{2\pi}{3}$,
\eqb
a_3+\bar{a}_3=-\frac{{\mathfrak r}_3(\phi)}{\sin \frac{\phi}{2}}.
\eqe

In this way, one can determine $Y_j^{(p,0)}$ successively. 
In particular, the  first non-trivial coefficients are especially simple, to be given 
explicitly by
\begin{align*}
2^{\frac{4}{3}} Y^{(1,0)}_1 &=
\frac{ 2\sqrt{3} \pi \alpha_1}
{ \Gamma(\frac{2}{3}+\frac{\phi}{2\pi})\Gamma(\frac{2}{3}-\frac{\phi}{2\pi})}\\
 2^{\frac{4}{3}} Y^{(1,0)}_2 &=
\frac{4 \sqrt{3}\alpha_1 \pi^2 }{
\Gamma(\frac{2}{3}+\frac{\phi}{2\pi})\Gamma(\frac{2}{3}-\frac{\phi}{2\pi})
\Gamma(\frac{1}{2}+\frac{\phi}{2\pi})\Gamma(\frac{1}{2}-\frac{\phi}{2\pi})
},
\end{align*}
with 
\begin{align*}
\alpha_1=\frac{1}{4\pi^{\frac{7}{6}}}\Gamma\( \frac{1}{6} \) \Gamma\( \frac{1}{3} \)
\( \frac{\Gamma(\frac{1}{4})}{\Gamma(\frac{3}{4})} \)^{-\frac{4}{3}}.
\end{align*}

As a check, one can compare the above result 
with the relations of $Y_j^{(p,q)}$ which follow from 
the Y-system (\ref{Ysystem1}), (\ref{Ysystem2}).
For example, the Y-system requires $Y_2^{(1,0)}/Y_1^{(1,0)} = 2 \cos(\phi/2)$, which indeed
agrees with (\ref{Y1p0}), (\ref{Y2p0}) and (\ref{a1}).
At the next order, the Y-system gives a linear relation among $Y_j^{(2,0)}$ and $(Y_j^{(1,0)})^2$.
This is also confirmed from the above result. Moreover, 
one finds that
$Y_j^{(0,1)} = 0$ $(j=1,2)$.   This means that, up to and including  $\cO(\ell^{4/3})$,
the Y-functions in 
the original TBA 
are fixed only by the information from the CFT limit. 

By recovering the phase $\varphi$ by $Y_j(\theta) \to Y_j(\theta - i \varphi)$, 
we finally obtain the expansion of the massive Y-functions,
\begin{align}\label{Yexpansion}
Y_j(\theta)=2Y_j^{(0,0)}+2Y_j^{(1,0)} \ell^{\frac{4}{3}}
 \cosh \( \frac{4(\theta-i\varphi)}{3} \)+\cO(\ell^{\frac{8}{3}}) \qquad (j=1,2).
\end{align}
For $\mu =1$, corresponding to the minimal surfaces in AdS$_{4}$, this reduces to the
expansion in \cite{Hatsuda:2012pb}. It is also in agreement with the expansion 
in \cite{Hatsuda:2010vr} with $\mu \neq 1$ determined numerically. 
Once the expansion of the Y-functions is found, one can immediately know 
the expansion of the T-functions through
the relation \eqref{eq:YT-rel-sc}.
The quantum Wronskian relation thus  provides a systematic and simple way to 
determine the coefficients $Y_j^{(p,0)}$ and $T_j^{(p,0)}$ .

\begin{figure}[t]
%\vspace{5ex}
 \begin{center}
   \begin{minipage}{0.4\hsize}
   \begin{center}
  \includegraphics[width=57mm]{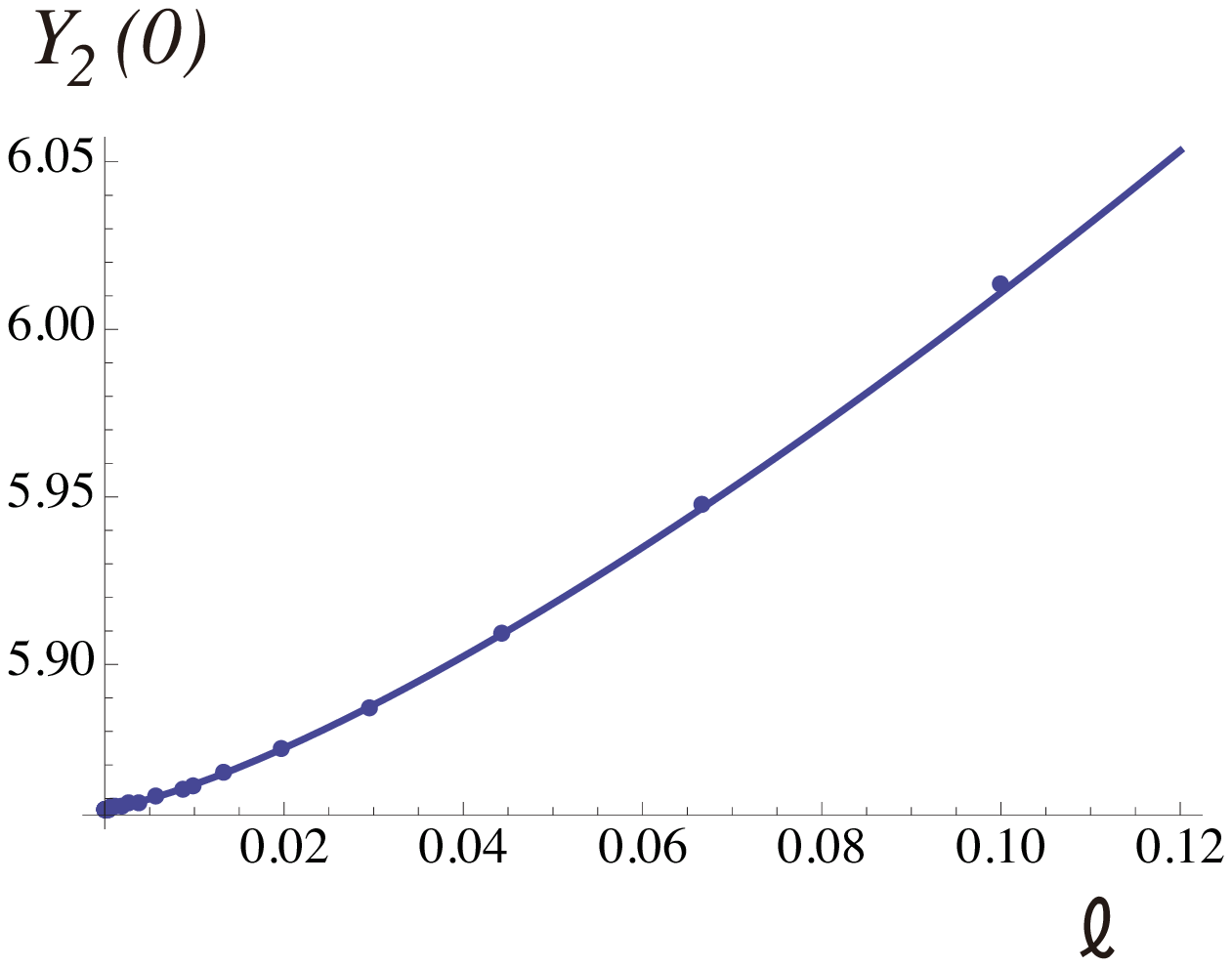}  \\ { \small (a)}
 \end{center}
 \end{minipage}
\hspace*{7ex}
\begin{minipage}{0.4\hsize}
%\vspace{5ex}
 \begin{center}
  \includegraphics[width=56mm]{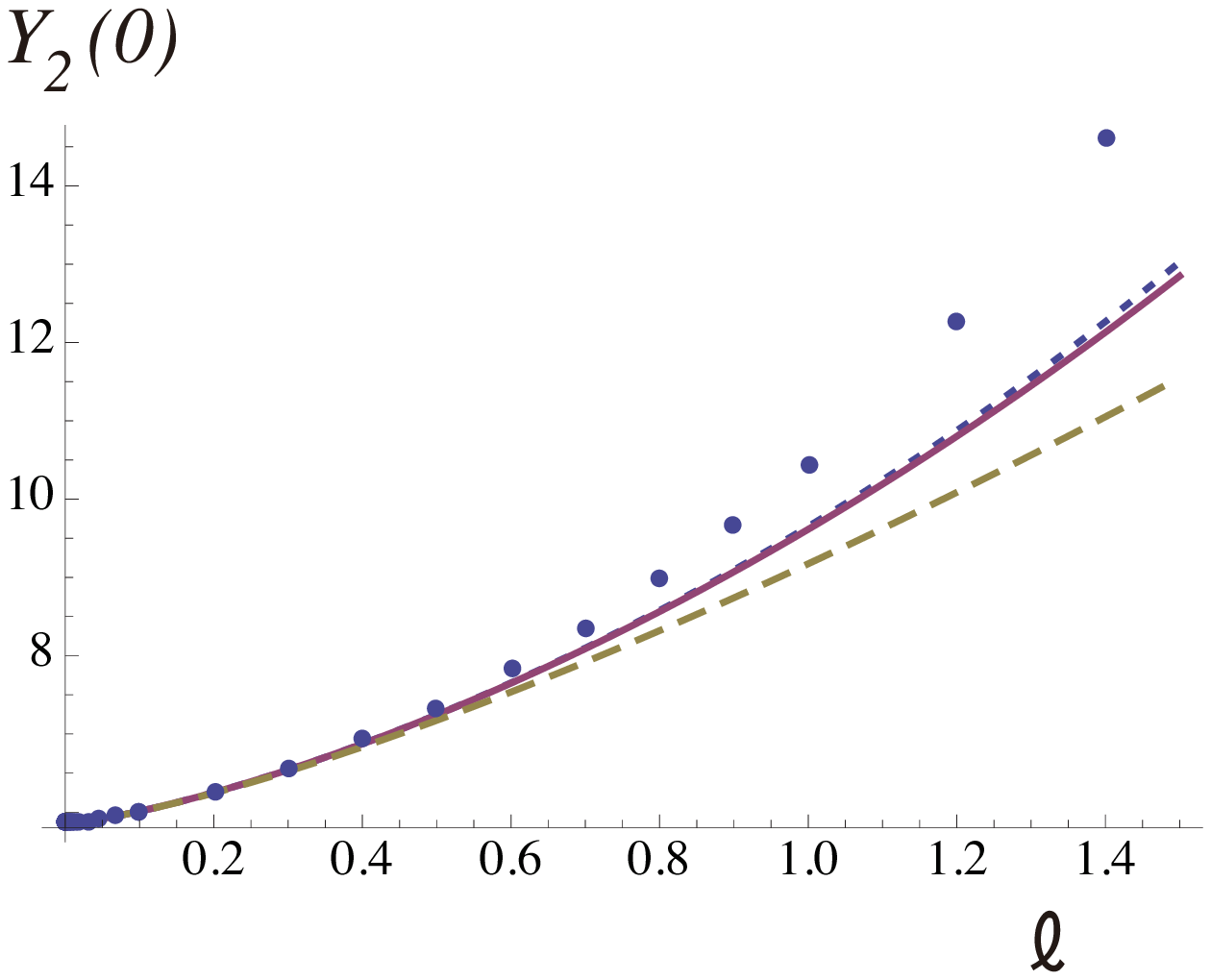}  \\ { \small (b)}
 \end{center}
 \end{minipage}
 \hfill
\caption{Plot of $Y_2(0)$ for $\varphi=-\pi/20$ and $\mu = 10$ as $\ell$ 
varies. In both (a) and (b), the points represent
numerical results. In (a), the solid line represents the leading expansion (\ref{Yexpansion}) 
evaluated at $\theta = 0$. (b) shows the expansions from the data in the CFT limit.
The broken ($--$), solid $(-)$ and dotted ($\cdots$)  lines represent 
$\sum_{p=0}^k 2 Y_2^{(p,0)}\cdot \ell^{4p \o 3} \cos \frac{4p}{3}\varphi  
$ with $k=1,2,3$, respectively.
The case with $k=1$ is equivalent to (\ref{Yexpansion}).}
 \label{fig:fig1}
\end{center}
\end{figure}

Figure \ref{fig:fig1} (a) shows a plot of $Y_2(0)$ as $\ell$ varies. 
As an example, the phase is fixed to be $\varphi = -\pi/20$ and the chemical potential 
to be $\mu = 10$.
We have chosen a real $\mu$ (imaginary $\phi$) 
so that we can compare our data against  the three-loop result 
in term of the multiple polylogarithms
\cite{Dixon:2013eka} , which is
discussed in the next section. 
Our expansions are valid also for real $\mu$. 
We find a good agreement between the numerical results and our analytic expansion.
A fit by a function $\sum_{k=0}^5 y_2^{(k)} l^{4k/3}$ for  $\calO(10^{-5}) < \ell< \calO(10^{-1}) $
can also reproduce $Y_2^{(0,0)}$ 
and  $Y_2^{(1,0)}$ with 12- and 8-digit accuracy, respectively. 
At $\calO(l^{8/3})$, the coefficient $Y_2^{(2,0)}$ explains about 44 per cent of 
$y_2^{(2)}$, whereas about 51 per cent are from $Y_2^{(0,2)}$, 
which is determined by and proportional
to $(Y_2^{(1,0)})^2$. The rest is carried by the undetermined $Y_2^{(1,1)}$.
At $\calO(l^{12/3})$, $Y_2^{(3,0)}$ explains about 16 per cent of 
$y_2^{(3)}$. 
 Figure \ref{fig:fig1} (b) shows a plot of the expansions of $Y_2(0)$ 
obtained from the analytic data  in the CFT limit,
i.e., $Y_2^{(p,0)}$, up to $p=1,2,3$, respectively. $\varphi$ and $\mu$ are the same as in (a). 
Up to $\ell \sim 1$, the CFT data  approximate $Y_2$ relatively well.
Combining them with the Y-system yields better  approximation.

%-------------------------------------------------------------------------------------
\section{Application to six-point amplitudes at strong coupling}
%-------------------------------------------------------------------------------------

Now, we apply the expansion of the Y-functions for small $\ell$  
to the six-point amplitudes or the null-polygonal Wilson loops dual to the amplitudes. 
Our evaluation based on the quantum Wronskian relation
allows us to analyze
the amplitudes/Wilson loops corresponding
to the minimal surfaces in AdS${}_5$ or $\mu \neq 1$.
In the small-$\ell$ or the UV limit, the Wilson loops become regular 
polygonal. The expansion thus gives the amplitudes/Wilson loops
under small deformations around the regular polygonal contour.

To evaluate the amplitudes, we first recall that the remainder function 
in (\ref{RemainderFn}) consists of three terms. One of the terms denoted by $A_{\rm free}$
is nothing but the free energy of the $\bbZ_4$-symmetric integrable model twisted by $\phi$.
Since this $\bbZ_4$-symmetric integrable model reduces, in the UV limit, to
the twisted $\bbZ_4$-parafermion 
 CFT, the free energy $A_{\rm free}$
is expanded by the bulk conformal perturbation theory 
\cite{Hatsuda:2010vr}:
\eqb
 A_{\rm free} \Eqn{=} \frac{\pi}{6} \Bigl( 1-  \frac{9\phi^2}{2\pi^2}\Bigr) -|Z|^2
 \nn \\
 && \quad  + \ 2 \pi^{1/3} \kappa_4^2 \gamma\Bigl(\frac{1}{3} \Bigr) 
\gamma\Bigl(\frac{1}{3} +\frac{\phi}{2\pi}\Bigr) 
\gamma\Bigl(\frac{1}{3} -\frac{\phi}{2\pi}\Bigr) \, |Z|^{\frac{8}{3}}
+ {\cal O}(|Z|^{\frac{16}{3}}) \comma
\eqe
where 
\be
 \kappa_4 =  \frac{1}{2\pi}  \gamma^{\frac{1}{2}}\Bigl(\frac{1}{6}\Bigr)
 \Bigl[\sqrt{\pi}\gamma\Bigl(\frac{3}{4} \Bigr)
  \Bigr]^{\frac{4}{3}} \comma
\ee
and $\gamma(z) = \Gamma(z)/\Gamma(1-z)$. The second term cancels $A_{\rm period}$.

The expansion of  the remaining term $\Delta A_{\rm BDS}$
is derived by utilizing the results in 
(\ref{Yspecial}), (\ref{crossratios}) and (\ref{RemainderFn}),
\be 
  \Delta A_{\rm BDS} 
 = -  \frac{3}{4} \Li_2\bigl(1- U_0 \bigr) 
 +  \frac{3\times  2^{\frac{2}{3}}(U_0-1 +\log U_0)}{ U_0(U_0-1)^2} (Y_2^{(1,0)})^2  
 |Z|^{\frac{8}{3}}
 +  {\cal O}(|Z|^{\frac{16}{3}}) \period
\ee
Here, $U_0$ is the value of $U_k $ in the UV limit,
\be
   U_0 = 1+ 2Y_2^{(0,0)} = 4 \cos^2\bigl({\phi}/{2} \bigr) \comma
\ee  
common to all $ k=1,2,3$.
We note that 
$\Delta A_{\rm BDS}$ 
is expanded in powers of $Z^{4/3} = |Z|^{4/3}e^{i \frac{4}{3}\varphi}$ and its
complex conjugate, but the $\varphi$-dependence remains only at $\calO (|Z|^{8n/3})$ with 
$n \in 3 \bbZ$. This is a consequence of 
the $\bbZ_6$-symmetry
of the six-point amplitudes $ \theta \to \theta + {\pi i}/{4}$ or  $ \varphi \to \varphi + {\pi}/{4}$,
which corresponds to cyclically  renaming the cusp points $x_a \to x_{a+1}$.
This symmetry also explains the absence of the $\calO(|Z|^{4/3})$ term.

Combining the above results, we obtain the UV 
expansion of the remainder function,
\be\label{UVExpansionR}
  R = \sum_{k=0}^\infty 
 r^{(k)}(\varphi,\phi) \, \ell^{\frac{8}{3} k} \comma
\ee
where 
\eqb\label{R01}
 r^{(0)}
  \Eqn{=}  -\frac{\pi}{6} + \frac{3}{4\pi} \phi^2  - \frac{3}{4} \Li_2 \bigl(1- U_0 \bigr) \comma \nn \\
 r^{(1)} 
 \Eqn{=}  \frac{3 \kappa_4^2}{32 (2\pi)^{\frac{2}{3}}} 
 \Bigl[ \Bigl(1-\frac{8\sqrt{3}}{9}\Bigr)(1- U_0) - \log U_{0} \Bigr] 
 \cdot B^2\Bigl(\frac{1}{3} +  \frac{\phi}{2\pi}, \frac{1}{3} - \frac{\phi}{2\pi}\Bigr) \comma
\eqe
and
$B(x,y)$ is the  beta function, $ \Gamma(x)\Gamma(y)/\Gamma(x+y)$.  
 The  expansion of the Y-functions (\ref{Yexpansion}) also yields the cross-ratios for small $\ell$, 
\be\label{UkExp}
 U_k = 
    U_0
  +2Y_2^{(1,0)} \ell^{\frac{4}{3}} \cos \left[\frac{4}{3} \Bigl( \frac{2k+1}{4}  \pi - \varphi \Bigr) \right]
 +\cO(\ell^{\frac{8}{3}})
 \period
\ee
Inverting this relation, one can express the parameters in the TBA equations by the cross-ratios
\cite{Hatsuda:2010vr}, 
\eqb
     \cos^2 \frac{\phi}{2} \Eqn{=} \frac{1}{12}(U_1+U_2+U_3), \nn \\
    \tan \frac{4}{3} \varphi \Eqn{=} 
    \frac{\sqrt{3}(U_2-U_3)}{2U_1 -U_2 -U_3},\\
    \ell^{\frac{4}{3}} \Eqn{=} 
    \frac{-2U_1 +U_2 +U_3}{6 Y_2^{(1,0)} \cos \frac{4}{3} \varphi},  \nn
\eqe
which is valid up to $\cO(\ell^{\frac{8}{3}})$. 

\begin{figure}[t]
\begin{center}
\begin{minipage}{0.4\hsize}
 \begin{center}
  \includegraphics[width=56mm]{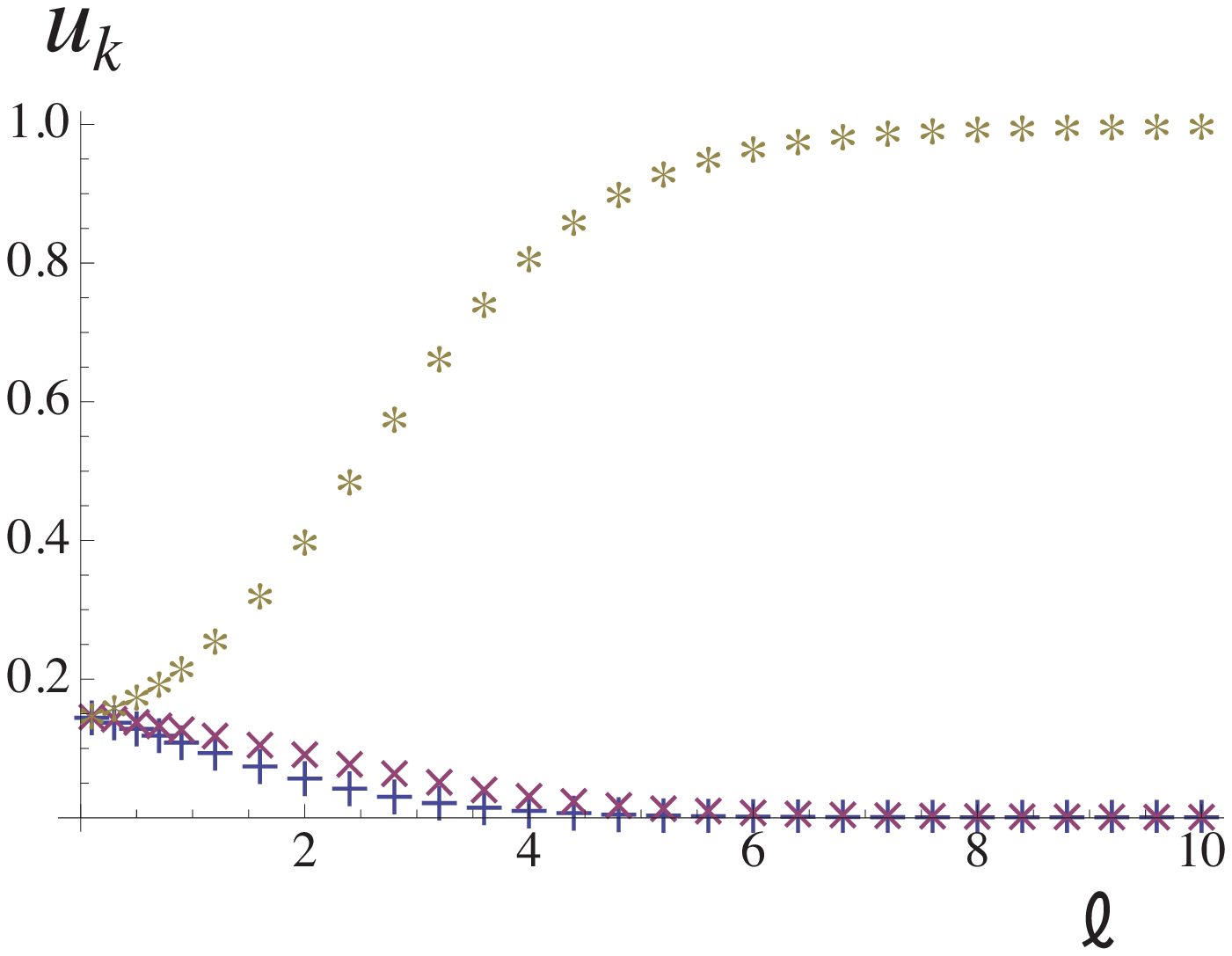}  \\  { \small (a)}
 \end{center}
\end{minipage}
\hspace*{7ex}
\begin{minipage}{0.4\hsize}
 \begin{center}
  \includegraphics[width=56mm]{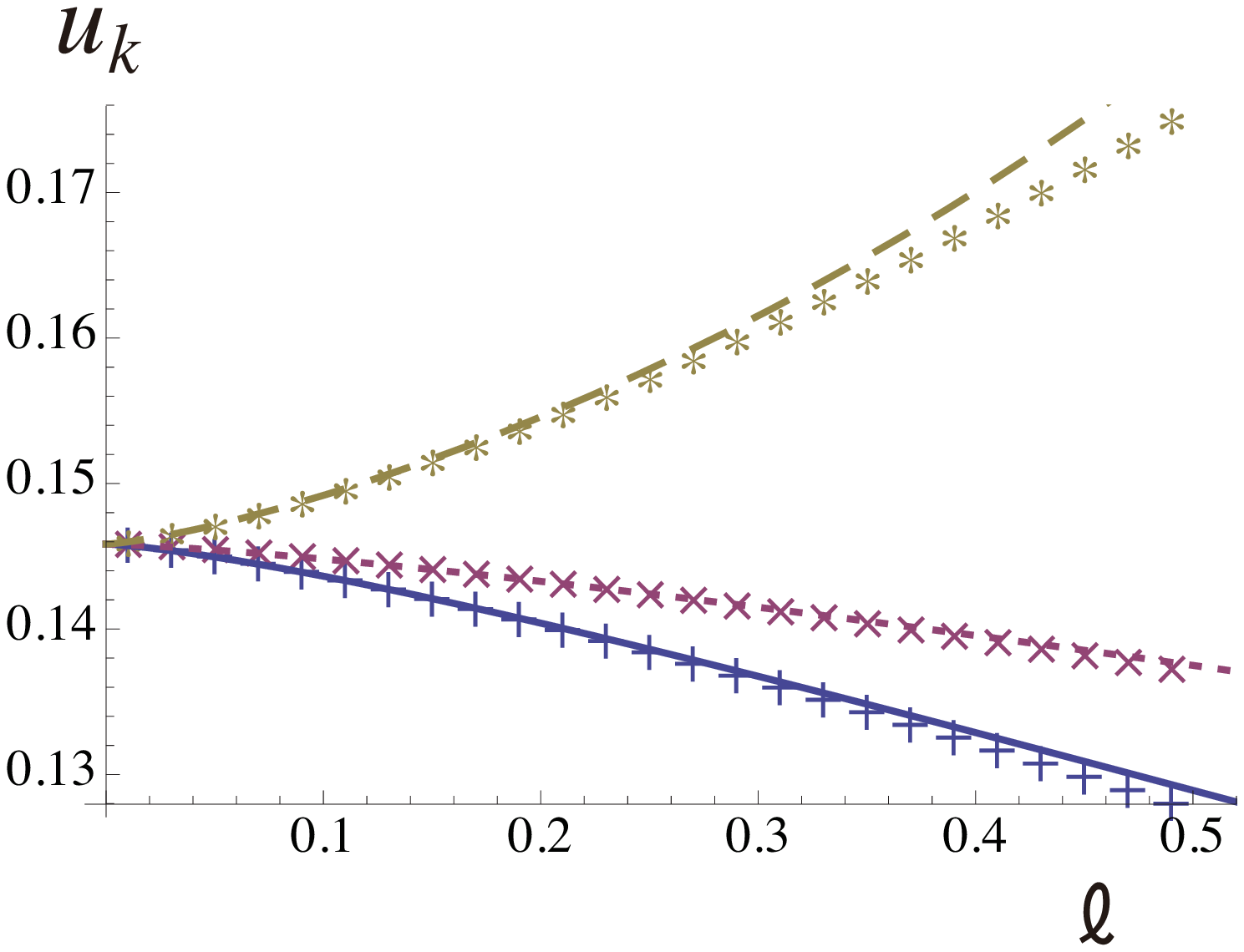}  \\ { \small (b)}
 \end{center}
 \end{minipage}
 \hfill
\caption{ (a) Trajectories of $u_k = 1/U_k$ as $\ell$ varies with $\varphi = - \pi/20$ 
and $ \mu = 10$. Points denoted by 
$ \ast, +, \times$ 
represent $u_1, u_2, u_3$, respectively. (b) The same trajectories for small $\ell$ (points)
and the expansion from  (\ref{UkExp}). 
 The broken ($--$), solid $(-)$ and dotted ($\cdots$)   lines represent
$u_1, u_2, u_3$, respectively. }
 \label{fig:fig2}
\end{center}
\end{figure}

\begin{figure}[t]
\begin{center}
\begin{minipage}{0.4\hsize}
 \begin{center}
  \includegraphics[width=60mm]{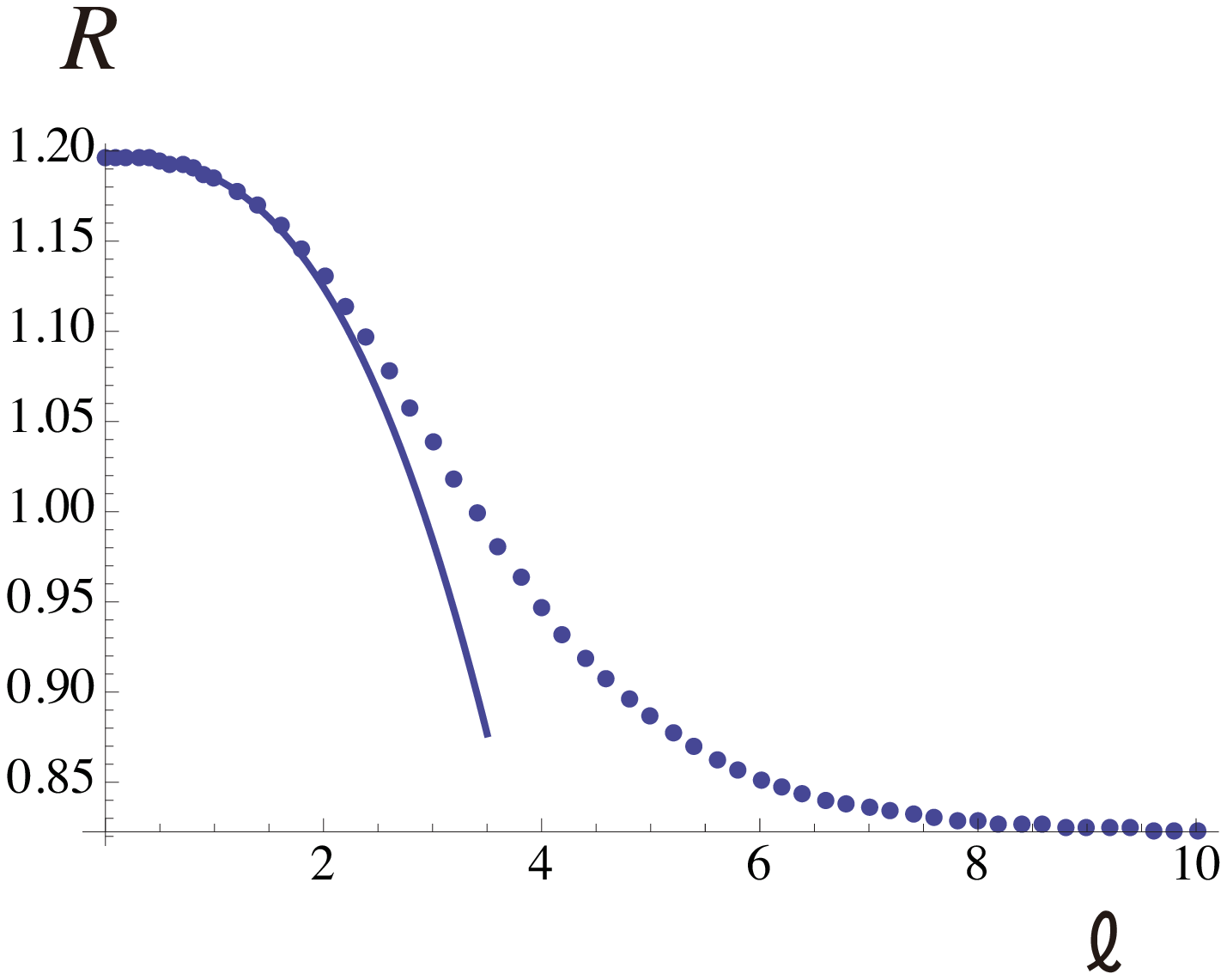}  \\  { \small (a)}
 \end{center}
\end{minipage}
\hspace*{7ex}
\begin{minipage}{0.4\hsize}
 \begin{center}
  \includegraphics[width=56mm]{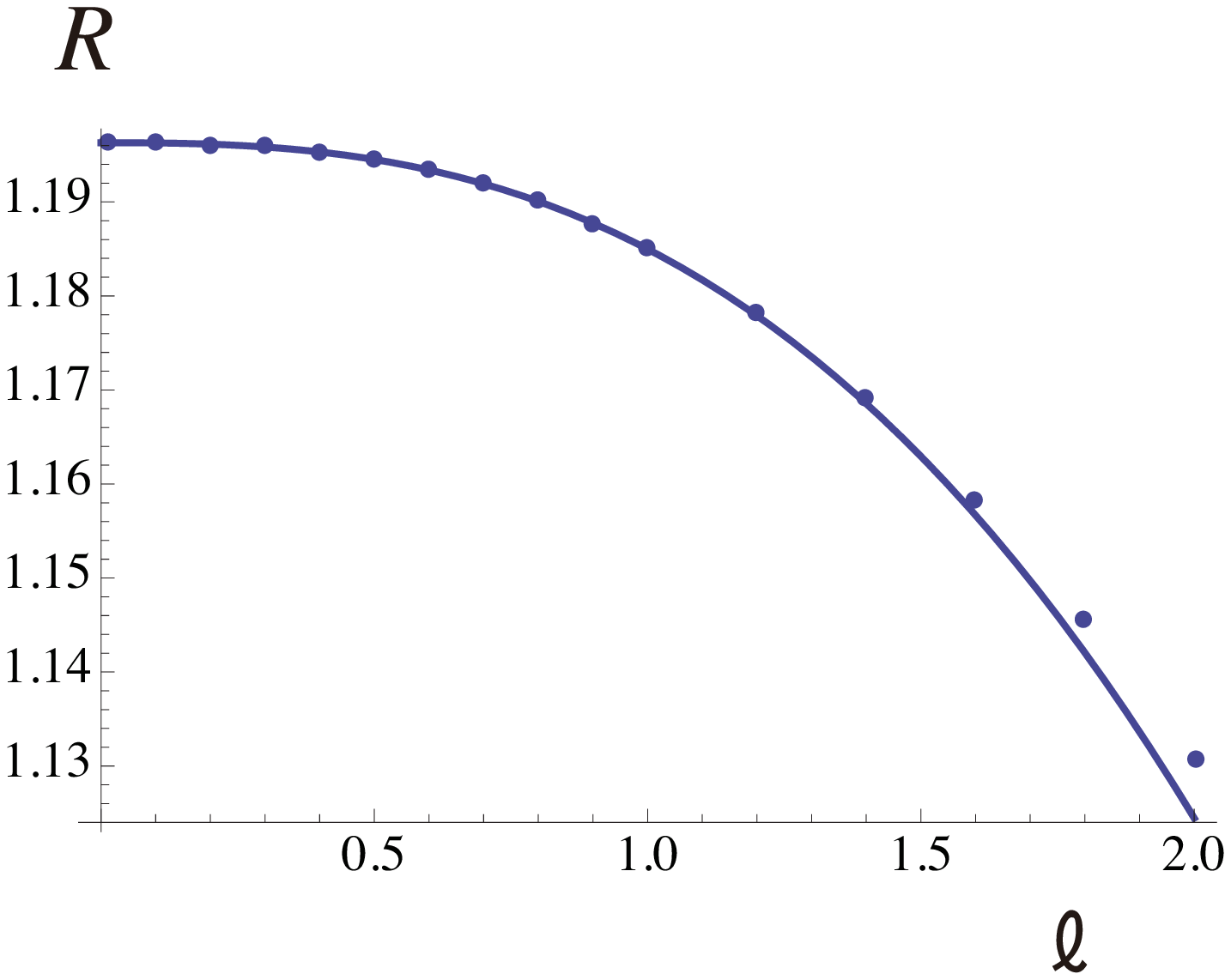}  \\ { \small (b)}
 \end{center}
 \end{minipage}
 \hfill
\caption{ (a) Six-point reminder function at strong coupling as $\ell$ varies 
with $\varphi = - \pi/20$ and $ \mu = 10$.
The points represent the numerical results, whereas the solid line represents our expansion  
(\ref{UVExpansionR})  with  (\ref{R01}). (b) The same plot for small $\ell$. }
 \label{fig:fig3}
\end{center}
\end{figure}

Our formulas can be checked numerically. Figure \ref{fig:fig2}  
(a) shows   the trajectories of $u_k =1/U_k$ as $\ell$ varies with 
$\varphi$ and $\mu = e^{\frac{3}{2}i\phi}$ fixed to be $-\pi/20$ and $10$, respectively. 
They are obtained by 
solving the TBA equations and evaluating $A_{\rm free}$ numerically.
Figure \ref{fig:fig2} (b) is a comparison of the same trajectories for small $\ell$ and the expansion 
of $u_k$  obtained from  (\ref{UkExp}).
Figure \ref{fig:fig3} (a) shows the remainder function at strong coupling 
$R$ for the cross-ratios $u_k$ given in 
Figure \ref{fig:fig2}. The points represent the numerical results, 
whereas the solid line represents our expansion  
$R = r^{(0)}+ r^{(1)} \ell^{8/3}$. 
Figure \ref{fig:fig3} (b) is the same plot for small $\ell$.
From Figure \ref{fig:fig2} and \ref{fig:fig3}, we find again a good agreement between 
the numerical results and
our analytic expansions for small $\ell$.

%-------------------------------------------------------------------------------------
\section{Comparison with perturbative results}
%-------------------------------------------------------------------------------------

In \cite{Brandhuber:2009da,Hatsuda:2011ke,Hatsuda:2011jn,Hatsuda:2012pb}, the  
remainder functions of the strong-coupling amplitudes corresponding to the minimal surfaces 
in AdS${}_3$ and AdS${}_4$ were compared with the two-loop results. It was found there that 
after an appropriate normalization/rescaling they are close to each other. 
In this section, we compare the six-point remainder function at strong  
with those at two \cite{DelDuca:2010zg,Goncharov:2010jf},  
three \cite{Dixon:2013eka}   and four \cite{Dixon:2014voa} loops.

For this purpose, we normalize/rescale the remainder function 
\cite{Brandhuber:2009da} so that 
it vanishes at $\ell=0$ and approaches $-1$ for large $\ell$: 
\be
 \bar{R}^{\rm strong}  = {R^{\rm strong} - R^{\rm strong}_{\rm UV} \over
  R^{\rm strong}_{\rm  UV} - R^{\rm strong}_{\rm  IR}} \comma
\ee
where we have introduced the notation $R_6^{\rm strong} := R$.
$R^{\rm strong}_{\rm UV}$ ($R^{\rm strong}_{\rm  IR}$) is the value at $\ell=0$ 
($\ell= \infty$) along the trajectory 
in the space of the
cross-ratios parametrized by $\ell$ with $\varphi, \mu$ fixed. 
For the remainder function at $L$ loops appearing in the perturbative expansion
$R_6 = \sum \lambda^L R^{(L)}$, one can also defined the rescaled reminder functions
$\bar{R}^{(L)}$ similarly.

At strong coupling, the UV value is read off from the expansion in (\ref{UVExpansionR}), 
$R^{\rm  strong}_{\rm UV} = r^{(0)}$.
To find the IR value $R^{\rm strong}_{\rm IR}$, we use the large-$\ell$
values of the cross-ratios 
$U_k \to (1, e^{\sqrt{2} \ell \cos(\frac{\pi}{4}+\varphi)}, e^{\sqrt{2} \ell \cos(\frac{\pi}{4}-\varphi)})$, 
which are 
found from the Y-system (\ref{Ysystem1}), (\ref{Ysystem2}) and 
the asymptotic behavior $Y_2(\theta) \to 2\sqrt{2} (\bar{Z} e^{\theta} +  Z e^{-\theta})$
for large $\ell$ with $|\Im \theta - \varphi | < \pi/2 $.
Since $A_{\rm free} \to 0$ and $A_{\rm period}$ cancels the leading term from 
$\Delta A_{\rm BDS}$ 
we are left with $R^{\rm strong}_{\rm IR} = {\pi^2}/{12} $.
On the perturbative side,  
 $R_{\rm UV}^{(L)} $  are obtained from  the cross-ratios $U_0$ in the UV limit,
whereas the IR values just vanish $R_{\rm IR}^{(L)} = 0$.

We evaluate these rescaled remainder functions for the cross-ratios $u_k (=1/U_k)$ 
given in 
Figure \ref{fig:fig2}, which are parametrized by  $\ell$  with $\varphi = - \pi/20, \mu = 10$.
At strong coupling, it is readily read off from the results in the previous section.
At two loops, we use the simple analytic expression given in \cite{Goncharov:2010jf}, whereas
at three loops we use the expression  given in 
\cite{Dixon:2013eka} and evaluate it by using the C++ library GiNaC 
\cite{Bauer:2000cp,Vollinga:2004sn}.
The direct three-loop expression in terms of the multiple polylogarithms 
is given for the parameter region 
with real $\mu$, which corresponds to the (2,2) signature of the four-dimensional space-time. 
In order to use this expression, we have chosen a real $\mu$. 
At four loops, it  can be evaluated from
the analytic expression in 
\cite{Dixon:2014voa}.\footnote{The four-loop data used here 
were provided to us by Lance Dixon,
James Drummond, Claude Duhr and Jeffrey Pennington.
We would like to thank them for providing these data.
}

\begin{figure}[t]
%\hspace*{5ex}
\begin{center}
\begin{minipage}{0.4\hsize}
 \begin{center}
  \includegraphics[width=62mm]{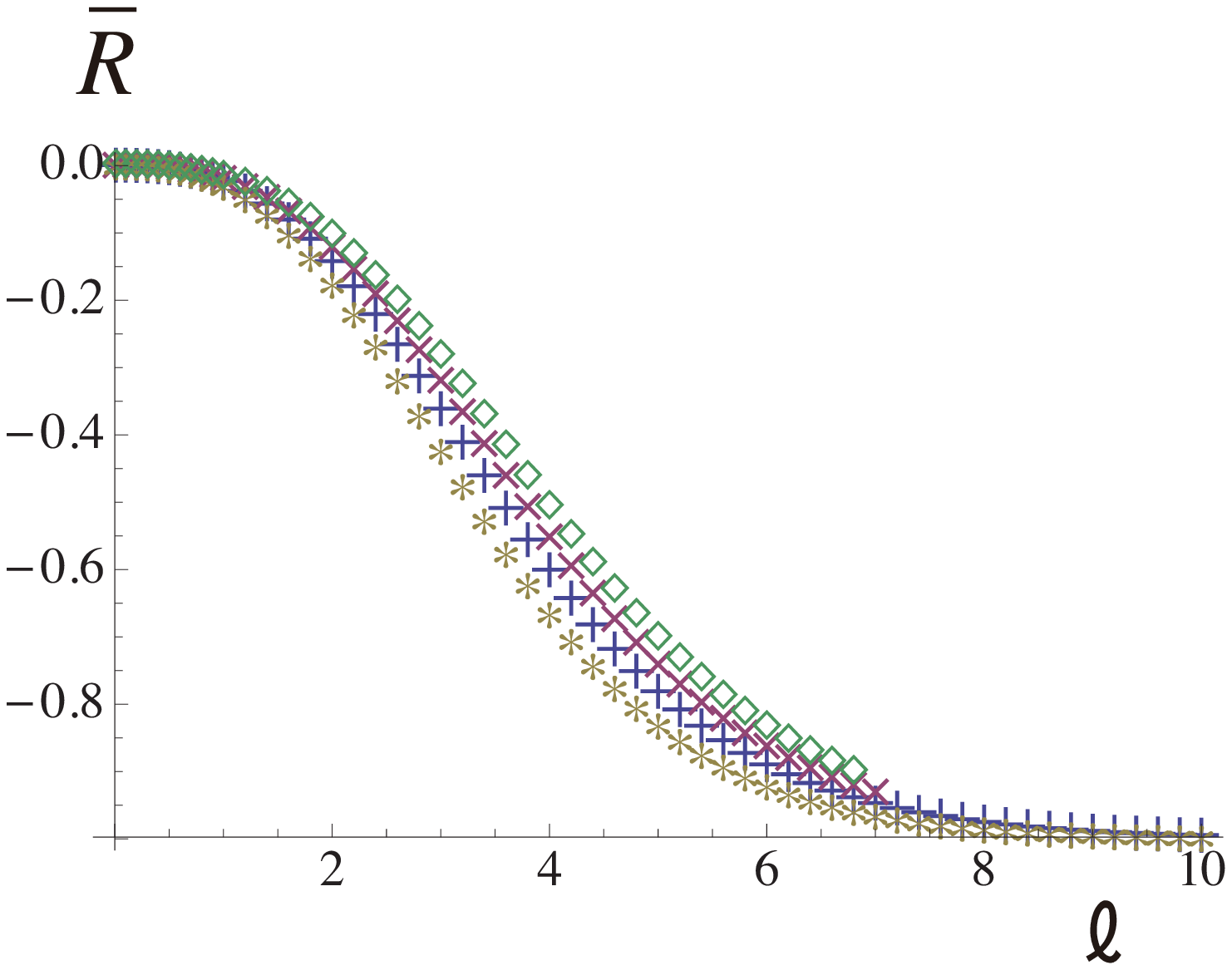}  \\  { \small (a)}
 \end{center}
\end{minipage}
\hspace*{7ex}
\begin{minipage}{0.4\hsize}
%\vspace{5ex}
 \begin{center}
  \includegraphics[width=57mm]{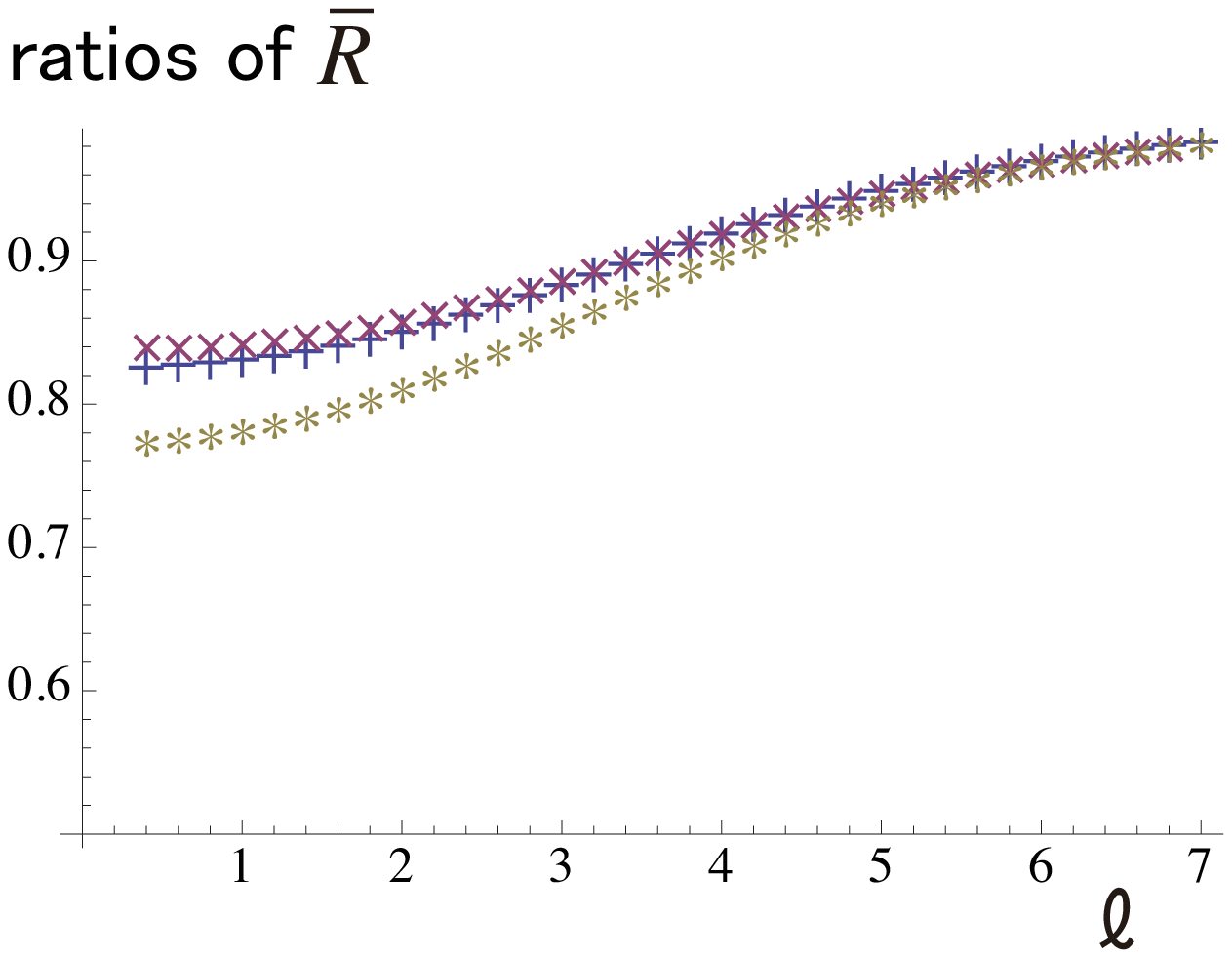}  \\ { \small (b)}
 \end{center}
 \end{minipage}
 \hfill
\caption{(a) Rescaled remainder functions at two loops ($+$), 
three loops $(\times)$, four loops ($\diamond$) and strong coupling
$(\ast)$ for the cross-ratios given in Figure \ref{fig:fig2} 
with $\varphi = - \pi/20$ and $ \mu = 10$. 
 (b) 
 Ratios of the rescaled remainder functions. Points denoted by $+, \times, \ast $ represent
$\bar{R}^{(3)}/\bar{R}^{(2)}$, $\bar{R}^{(4)}/\bar{R}^{(3)}$ 
and $\bar{R}^{(2)}/\bar{R}^{\rm strong}$, respectively.}
 \label{fig:fig4}
\end{center}
\end{figure}

Figure \ref{fig:fig4} (a) is a plot of the rescaled remainder functions 
at two, three and four loops and at strong coupling. 
Figure \ref{fig:fig4} (b) is a plot of the ratios of these rescaled 
remainder functions; $\bar{R}^{(3)}/\bar{R}^{(2)}$,  $\bar{R}^{(4)}/\bar{R}^{(3)}$
and $\bar{R}^{(2)}/\bar{R}^{\rm strong}$.
As $\ell$ increases, it becomes harder to 
evaluate $R^{(3)}$ and Figure \ref{fig:fig4} includes the data for $\ell\leq 7$. 
It also includes the four-loop data  for $\ell\leq 34/5$.
Some of the numerical values of the rescaled remainder functions plotted in Figure \ref{fig:fig4}
are listed in Table \ref{Table1}.
{}From these figures and the table,
we find that the rescaled remainder functions stay close to each other 
as $\ell$ varies, but the perturbative results gradually move away from those at strong coupling
as the number of loops increases.
Their ratios changes slowly along $\ell$, and accumulate to $1$ for large $\ell$ 
as assured by definition. The ratios of the perturbative results, 
$\bar{R}^{(3)}/\bar{R}^{(2)}$,  $\bar{R}^{(4)}/\bar{R}^{(3)}$, are very similar.
These are in accord with the observations in 
\cite{Brandhuber:2009da,Hatsuda:2011ke,Hatsuda:2011jn,Hatsuda:2012pb}
mentioned above and those in  \cite{Dixon:2013eka,Dixon:2014voa} that the ratios 
of the remainder functions at two, three and four loops are relatively constant 
for large ranges of the cross-ratios.

 As $\mu$ increases, we have observed that
 $\bar{R}_6^{(2)}$, $\bar{R}_6^{(3)}$ and $\bar{R}_6^{\rm strong}$ change 
in a similar manner: they tend to 
start decreasing for larger $\ell$.
Although their differences increase, they are still kept relatively close to each other.
The dependence on $\mu$ seems very weak. 
For example, even for $\mu = 10^6$, the ratios are still 
of $\calO(1)$.
At the order of the expansion in (\ref{UVExpansionR}) and (\ref{R01}), the $\varphi$-dependence 
does not appear. 
Although it does appear at higher orders, the behavior of the rescaled
remainder functions is still qualitatively similar.

\begin{table}[t]
 \centering 
\begin{tabular}{ |c || l | l | l | l | l | l | l |}
\hline
  $\ell$ & \qquad \, 1/5 &  \quad \ \ 1  &  \quad \ 3  &  \quad \ 5  &  \hspace{0.7ex} 34/5  &   
\quad \ 9 &   \quad  10    \\
\hline
\hline
 $\bar{R}^{(2)}$ & $-3.202 \times 10^{-4}$ &  -0.02351  & -0.3618  & -0.7814  
& -0.9388  & -0.9890  &  -0.9951\\
 \hline
 $\bar{R}^{(3)}$ & $ -2.626  \times 10^{-4}$ &  -0.01953  & -0.3194  & -0.7411  
 & -0.9204  &   \quad \ --  \quad
  &   \quad  \ --  \quad  \\
 \hline
 $\bar{R}^{(4)}$ & $ -2.214  \times 10^{-4}$ &  -0.01643  & -0.2827  & -0.7014  
 & -0.9004  &   \quad \ --  \quad
  &   \quad  \ --  \quad  \\ 
 \hline
 $\bar{R}^{\rm strong}$ &  $ -4.145  \times 10^{-4}$ &  -0.03007  & -0.4226 
  & -0.8304  & -0.9584  & -0.9935 & -0.9973 \\
\hline
\end{tabular}
\caption{Samples of the numerical values of the rescaled remainder functions plotted in Figure \ref{fig:fig4}.
}
\label{Table1}
\end{table}

\section{Conclusions}

In this paper 
we studied six-point gluon scattering amplitudes in ${\cal N}=4$
super Yang-Mills theory at strong coupling by using the AdS/CFT correspondence.
The area 
of the corresponding null-polygonal minimal surface in AdS${}_5$  is evaluated by 
solving the T-/Y-system for the $\mathbb{Z}_4$-symmetric integrable model with a twist parameter.  
The  leading expansion to the remainder function is determined 
around the UV limit explicitly. 
We compared this result with the recent perturbative calculations,  
to find that the rescaled remainder functions are close to each other 
along the trajectories parametrized by the scale parameter. 

Our results at the leading order relied on the quantum Wronskian relation, which determines 
the expansion of the T-/Y-functions around the CFT (massless) limit.
For higher order terms, 
one needs to study the massive TBA system intrinsically.
As in \cite{Hatsuda:2011ke, Hatsuda:2011jn, Hatsuda:2012pb}, 
the boundary CPT would be a way toward this direction
based on the relation between the T-/Y-functions and
the $g$-function \cite{Bazhanov:1994ft,Dorey:1999cj,Dorey:2005ak}.
Through the T-/Y-functions, one could also study the quantum Wronskian 
relation from a different perspective by using the boundary CPT.
However, it is yet to be figured out 
how to incorporate  
the twist of the perturbed $\mathbb{Z}_4$-parafermion  theory in this framework. 
 
 Alternatively, the  quantum  Wronskian relation exists even in the massive case \cite{LZ:2010}.
 The more involved analyticity, however,  defies the analytical determination of the moments
 of ${\bf A}$, $\bar{\bf A}$. Along \cite{Bazhanov:1996dr,Bazhanov:1994ft}, 
 the massive TBA systems have also been analyzed  in 
 \cite{Bazhanov:1996aq, Fioravanti:2003kx}.
 Hopefully one can detour the difficulties, and obtain the systematic higher expansions.

Finally, given the recent developments on the finite-coupling amplitudes around the collinear limit 
\cite{Basso:2013vsa,Basso:2013aha,Basso:2014koa}, it would be worthwhile to explore
the extrapolation to the finite coupling also around the regular-polygonal limit.
We hope to come back to these issues in near future.

%-------------------------------------------------------------------------------------
\subsection*{Acknowledgements} 
%-------------------------------------------------------------------------------------
We would like to thank Lance Dixon,
James Drummond, Claude Duhr and Jeffrey Pennington
for providing to us the four-loop data of the remainder function.
We would also like to thank Lance Dixon and Takahiro Ueda for useful correspondences. 
The work of K.~I., Y.~S. and J.~S. is supported in part by 
 JSPS Grant-in-Aid for Scientific Research (C) No. 23540290, 24540248 and 24540399.
The work of K.~I. and Y.~S. is also supported in part by JSPS
Japan-Hungary Research Cooperative Program.

\appendix
%-------------------------------------------------------------------------------------
\section{Non-Linear Integral Equation }
%-------------------------------------------------------------------------------------

We supplement a treatment on the Bethe ansatz equations 
based on  suitably chosen   auxiliary functions.
The approach utilizes  Non Linear Integral Equations  satisfied by them\cite{KBP, Destri:1992qk},
and thus is referred to as the  NLIE approach.%
\footnote{It is also referred to as the DDV approach in the context of
integrable field theories.}
Here we choose auxiliary functions    as 
\begin{align*}
\mathfrak{a}(x) &= 
\frac{\Phi(x-i\frac{\gamma}{2}) {\mathbf Q}(x+i\gamma) }{\Phi(x+i\frac{\gamma}{2})
 {\mathbf Q}(x-i\gamma)    } ,&
	 \mathfrak{A}(x) &= 1+\mathfrak{a}(x) ,  \\
\bar{\mathfrak{a}}(x) &=  
\frac{\Phi(x+i\frac{\gamma}{2}) {\mathbf Q}(x-i\gamma)    }{\Phi(x-i\frac{\gamma}{2})
{\mathbf Q}(x+i\gamma) } ,&
 \bar{\mathfrak{A}}(x) &= 1+\bar{\mathfrak{a}}(x).
\end{align*}
The Bethe ansatz equations are equivalent to
\begin{equation*}
\mathfrak{a}(x_j) =-1.
\end{equation*}
Numerically,  one observes the following properties of the finite size system  in the ground state:
\begin{enumerate}
\item  $|\mathfrak{a}(x) | <1\,  (|\bar{\mathfrak{a}}(x) | <1)$ in the upper (lower)
half plane.   
\item In a  narrow strip including
the real axis,  the zeros of  $\mathfrak{A}(x)\, (\bar{\mathfrak{A}}(x))$ 
are located only on the real axis and they coincide with
the Bethe ansatz roots $\{x_j\}$. 
\item 
The extended branch cut function, $\frac{1}{i} \ln \mathfrak{a}(x)$, 
is an increasing function of real $x$.
\end{enumerate}

These properties  are enough in  deriving the NLIE. 
After the scaling and the conformal  limit,  for  $\Im \theta$ positive small, it is explicitly given by
\begin{align}
\ln  \mathfrak{a}(\theta) &=d_a(\theta) + \int_{-\infty+i 0}^{\infty+i 0}  F (\theta-\theta') 
\ln \mathfrak{A}(\theta') \frac{d\theta'}{2\pi}
 -\int_{-\infty-i 0}^{\infty-i 0}   F(\theta-\theta') \ln \bar{\mathfrak{A}}(\theta')
  \frac{d\theta'}{2\pi},   \nonumber \\
d_a(\theta)&= i {\rm e}^{\theta}  - i \frac{3\phi}{2}  , \qquad 
F(\theta) = -\int_{-\infty}^{\infty} \frac{{\rm e}^{ik\theta}}{2 \cosh \frac{\pi}{2} k} dk,  \label{nliesc} 
\end{align}
where we use the same symbols $\mathfrak{a}$ etc for functions of 
$\theta (=\frac{3x}{2})$  by abuse of notation.

We are interested in  $\phi \gg 1 $ behavior of $\ln {\bf A}(\theta)$.
The driving term $d_a$ in the above suggests that  there exists $B$ of order $\ln \phi$ such that 
the smallest root  $\theta_1$  is greater than $B$.  
We  assume that  
$ |\ln \mathfrak{A}(\theta)| \ll 1$ and  $ |\ln \bar{\mathfrak{A}}(\theta)| \ll 1$  if
$\theta>\theta_1$.
Introduce 
$$
g(\theta) = \ln  \frac{ \mathfrak{A}(\theta+B+i\epsilon) }{ \bar{\mathfrak{A}}(\theta+B-i\epsilon) }.
$$
Then for $\theta\le 0$, the NLIE is approximated by  the following Wiener-Hopf type equation 
\cite{Bazhanov:1996dr}, 
\eqb
g(\theta) =d_a(\theta +B) + \int_{-\infty}^{0} F(\theta-\theta') g(\theta')  \frac{d\theta'}{2\pi}.
\label{WH}
\eqe

It is convenient to deal with the equation in the Fourier space,
$$
\widehat{g}(\omega)=  \int_{-\infty}^{\infty}  g(\theta)  {\rm e}^{-i\omega \theta} \frac{d\theta}{2\pi},
\quad
g(\theta)=  \int_{-\infty}^{\infty} \widehat{g}(\omega)   {\rm e}^{i\omega \theta} d\omega.
$$
The standard recipe is to introduce the factorized Kernel,
$$
G_+(\omega)G_-(\omega) =\Bigl( 1-\widehat{F}(\omega) \Bigr)^{-1} ,\qquad 
G_-(\omega) =G_+(-\omega)
$$
such that 
$$ 
\lim_{\omega \rightarrow \infty    }     G_{\pm}(\omega) \rightarrow 1.
$$
Explicitly we choose 
\begin{align*}
G_+(\omega)&=\sqrt{\frac{2 \pi}{3}} 
\frac{\Gamma(1-\frac{3\omega}{4}i)}
{ \Gamma(\frac{1}{2}-\frac{\omega}{2}i)  \Gamma(1-\frac{\omega}{4} i)} 
{\rm e}^{i\alpha \omega}  ,
\\
\alpha&=\frac{1}{2} \Bigl( \ln \Bigl( \frac{3}{2}\Bigr)^{\frac{3}{2}}
 - \ln \Bigl( \frac{1}{2}\Bigr)^{\frac{1}{2}} \Bigr).
\end{align*}
Then the solution in the Fourier space is given as follows,
\begin{align*}
\widehat{g}(\omega) &=G_+(\omega) \mathcal{Q}_+(\omega) , \\
\mathcal{Q}_+(\omega) &=\frac{1}{2\pi}
\Bigl(
\frac{3 \phi}{2}  \frac{G_-(-i\epsilon) }{\omega+i\epsilon}-  \frac{ G_-(-i)}{\omega+i} {\rm e}^B
\Bigr).
\end{align*}
The comparison of the asymptotic behavior $\omega \rightarrow \infty$  of the 
Fourier transformation of (\ref{WH}) concludes $\mathcal{Q}_+(\infty)=0$.
This determines the
relation between parameters $B$ and $\phi$,
\begin{equation}
{\rm e}^B =\frac{\phi}{2\sqrt{\pi}}
 \frac{\Gamma(\frac{1}{4})}{   \Gamma(\frac{3}{4}) } {\rm e}^{\alpha}.
\label{Bsol}
\end{equation}
This makes the expression for $ \widehat{g}(\omega) $ simpler,
\begin{equation} \label{gsol}
\widehat{g}(\omega)= \frac{i \phi  \Gamma(1-\frac{3\omega}{4}i)}
{2\sqrt{\pi}  (\omega+i\epsilon)(\omega+i) 
 \Gamma(\frac{1}{2}-\frac{\omega}{2}i) \Gamma(1-\frac{\omega}{4} i)      } 
{\rm e}^{i\alpha \omega}. 
\end{equation}
We introduce  $\mathbb{A}_+(\theta) := \ln \mathbf{A}(\theta+B)$ for $\theta<0$.
A simple manipulation  leads to the following expression of  its
Fourier transformation $\widehat{\mathbb{A}}_+(\omega)$ in terms of $\widehat{g}(\omega)$,
\begin{equation}
\widehat{\mathbb{A}}_+(\omega) =      
 \frac{  \widehat{g}(\omega)   }{  
  4 \sinh \frac{\pi }{ 4 } \omega \cosh \frac{\pi}{2}\omega     } {\rm e}^{\frac{3\pi}{4} \omega } . 
\label{lnAsol}
\end{equation}

By substituting (\ref{gsol}) into (\ref{lnAsol}), one finds that 
$\omega = -\frac{4}{3} (n+1) i, n\in \mathbf{Z}_{\ge 0}$
are  the only relevant poles in the inverse Fourier transformation of 
$\widehat{\mathbb{A}}_+(\omega)$.
We thus finds for $\phi \gg 1$,
\begin{equation*}
\ln \mathbf{A}(\theta) =- \frac{\phi}{4 \sqrt{\pi}} 
\sum_{n \ge 1}  {\rm e}^{\frac{2\gamma}{\pi}n (\alpha+\theta-B)}
\frac{\Gamma( n(1-\frac{\gamma}{\pi})) \Gamma(\frac{1}{2}+ \frac{\gamma}{\pi}n)}
{(2\gamma n-\pi)n!} .
\end{equation*}
By comparing the above expansion with (\ref{Aexpansion}), 
making use of   the explicit  form of B  in (\ref{Bsol}), 
we have the following asymptotic behavior of  $a_n(\phi)$,
$$
a_n(\phi) \sim \frac{ \phi}{8 \pi^{\frac{3}{2}}n!}
\Gamma(\frac{n}{3})\Gamma(\frac{ 2n}{3}-\frac{1}{2} )
\Bigl(
\frac{\phi}{2\sqrt{\pi}} 
\frac{\Gamma( \frac{1}{4})}{ \Gamma(  \frac{3}{4})   }
\Bigr)^{-\frac{4}{3}n}.
$$
The explicit form of the coefficient $\alpha_n$ in (\ref{an:asymp}) is  then determined as
in  (\ref{alphanso1}).

%%%%%%%%%%%%%%%%%%%%%%%%%%%%%
%     references
%%%%%%%%%%%%%%%%%%%%%%%%%%%%%
%
\def\thebibliography#1{\list
{[\arabic{enumi}]}{\settowidth\labelwidth{[#1]}\leftmargin\labelwidth
 \advance\leftmargin\labelsep
 \usecounter{enumi}}
 \def\newblock{\hskip .11em plus .33em minus .07em}
 \sloppy\clubpenalty4000\widowpenalty4000
 \sfcode`\.=1000\relax}
\let\endthebibliography=\endlist
\vspace{3ex}
\begin{center}
{\bf References}
\end{center}
\par \vspace*{-2ex}

\end{document}